\documentclass[aps,twocolumn,floats,showpacs]{revtex4}

\usepackage{amsmath,amssymb,graphicx}
\usepackage{epsfig}
\usepackage{graphicx}
\usepackage{enumerate}
\newcommand{\beq}{\begin{equation}}
\newcommand{\eeq}{\end{equation}}

\newcommand{\bea}{\begin{eqnarray}}
\newcommand{\eea}{\end{eqnarray}}

\begin{document}

\title{Non-Equilibrium Quantum Dissipation}

\author{
Dvira Segal\footnote{Corresponding author. Current Address:
Chemical Physics Theory Group, Department of Chemistry, University
of Toronto, Ontario M5S 3H6, Canada. Email:
dsegal@chem.utoronto.ca} and David R. Reichman}
\affiliation{Department of Chemistry, Columbia University, 3000
Broadway, New York, NY 10027}

\author{Andrew J. Millis}
\affiliation{Department of Physics, Columbia University, 538 W 120th St., New York, NY 10027.}

\begin{abstract}
Dissipative processes in non-equilibrium many-body systems are
fundamentally different than their equilibrium counterparts. Such
processes are of great importance for the understanding of
relaxation in single molecule devices. As a detailed case study,
we investigate here a generic spin-fermion model, where a
two-level system couples to two metallic leads with different
chemical potentials.
We present results for the spin relaxation rate in the
nonadiabatic limit for an arbitrary coupling to the leads, using
both analytical and exact numerical methods. The non-equilibrium
dynamics is reflected by an exponential relaxation at long times
and via complex phase shifts, leading in some cases to an
"anti-orthogonality" effect. In the limit of strong system-lead
coupling at zero temperature we demonstrate the onset of a
Marcus-like Gaussian decay with {\it voltage difference}
activation. This is analogous to the equilibrium spin-boson model,
where at strong coupling and high temperatures the spin excitation
rate manifests temperature activated Gaussian behavior. We find
that there is no simple linear relationship between the role of
the temperature in the bosonic system and a voltage drop in a
non-equilibrium electronic case. The two models also differ by the
orthogonality-catastrophe factor existing in a fermionic system,
which modifies the resulting lineshapes.
Implications for current characteristics are discussed.
We demonstrate the violation of
pair-wise Coulomb gas behavior for strong coupling to the leads.
The results presented in this paper form the basis of an exact, non-perturbative
description of steady-state quantum dissipative systems.
\end{abstract}

 \maketitle

\section{Introduction}

Over the past several decades tremendous effort has been put forth
to understand the dynamics of a small quantum entity coupled to a
thermal bath \cite{WeissBook}. Important problems that can be
distilled to this form include the interaction between localized
magnetic impurities and itinerant electrons (the Kondo problem)
\cite{Kondo,Kondobook}, electron transfer in aqueous environments
\cite{Ulstrup}, and proton tunneling in biomolecules
\cite{Bell,Proton}.
The study of such quantum dissipative systems cuts across traditional disciplines and
impacts fields from biology to quantum information theory \cite{WeissBook}.

Our detailed understanding of quantum dissipative systems is essentially
confined to problems which involve a single thermal reservoir \cite{Leggett}.
In this case, traditional measures of dynamical interest are
equilibrium correlation functions or simple measures of the decay
of one-time quantities when the initial condition is not one of
thermal equilibrium for the global system. While important for
the understanding of various experimental situations, this latter
form of non-equilibrium behavior is well understood and
generically takes the form of an asymptotic exponential decay
to the thermal equilibrium state or the ground state
(at zero temperature)  \cite{WeissBook,Leggett,Saleur}.

A less well understood type of non-equilibrium behavior may
manifest when a small quantum system is coupled to {\em more than
one reservoir}
\cite{Aditiphonon,Aditisemi,Aditispin,Aditiphase,Aditicoul,Zawadowski,Paaske03,Paaske04,Paaske,Kehrein,AndreiPRL,Andrei,
Kondononeq,KondoV}.
Here, the generic situation is one of a non-equilibrium steady
state, regardless of the initial preparation. Given the fact that
this multi-bath scenario is standard for prospective
single-molecule devices \cite{MolEl,NitzanRatner} as well as
more general problems, it is imperative to understand the
fundamental relaxation motifs that emerge in such nontrivial
non-equilibrium cases. Recent work raises the question of whether
standard tools borrowed from typical equilibrium quantum
dissipative systems are useful in the steady state non-equilibrium
case. For example, the simple equivalence between bosonic and
fermionic baths (as obtained via bosonization
\cite{Schotte,Giamarchi,Hakim}) is lost in the multi-bath case,
while mean-field approaches are fraught with danger due to the
fact that a voltage bias may assist tunneling even at
zero-temperature, rendering the meaning and stability of
Hartree-Fock minima unclear
\cite{Aditisemi,Alexandrov,Komnik,Nitzan-nano,Galperin}.

While several recent papers have taken up the task of describing
the steady state, non-equilibrium dynamics in different model
problems, our goal here is a first step towards a
{\em detailed and systematic} understanding of dissipative
relaxation in the simplest model problems resulting from coupling
a small quantum system to several baths, namely generalized
spin-boson models \cite{Aditiphonon,Aditisemi,Aditispin}.
It should be noted that the term ``spin-boson'' is a misnomer; the interesting and relevant case is
that of fermionic reservoirs, which dramatically differ from the
case of bosonic reservoirs when the system interacts with more
than one bath with different chemical potentials.
On the other hand, as in the standard spin-boson
model, it is the physics of the x-ray edge singularity
\cite{Nozieres,Ohtaka,LevitovM,Baranger} that forms the fundamental
building block of the description of dynamic observables.
Here, it is the recently studied {\it non-equilibrium} x-ray edge problem
\cite{Ng,Combescot,Braunecker0,Braunecker1,Muzy2,Braunecker2,Levitov} that lies at the
core of the relaxation behavior of standard correlation functions.
The more complex physics of the non-equilibrium edge behavior
allows for a richer range of dynamical behavior than in the
well-studied equilibrium case.

In this paper we will confine our discussion to calculations that are
perturbative in the bare tunneling matrix element of the system,
{\em but allow for arbitrarily strong coupling to the leads}.  We
will employ both analytical and numerical techniques to describe
the dynamics.  The numerical approach involves a computational
solution of the non-equilibrium x-ray edge problem that is {\it numerically exact}
on all relevant time scales.  This will allow us to describe the
full cross-over behavior from the regime where equilibrium effects
dominate, to that where the full non-equilibrium behavior (such as
bias-induced dephasing and complex phase shifts) is manifested.
This is crucial, since the full frequency dependence of relaxation
rates and generalized fluctuation-dissipation ratios depend on the
entire time history of the dynamics \cite{Aditispin}.

We will demonstrate that interesting behavior occurs in specific
parameter regimes that lead to anti-orthogonality effects and
bias-induced tunneling. In particular, the bias-induced tunneling
regime at {\em zero-temperature} may display a very broad Gaussian
decay of the polarization at strong system-leads coupling.  In
this regime, the relaxation behavior shows interesting
similarities to the usual {\em high-temperature} Marcus (or
semiclassical polaron) behavior \cite{Holstein,Mahan,Marcus}, with
potential bias playing the role of temperature, although crucial
differences exist that make these analogies imprecise. Lastly, we
investigate the crucial question of the accuracy of the pair-wise
Coulomb gas decomposition for non-equilibrium steady state
systems. We note that the methods discussed in this work form the
basis of a numerically exact path-integral description of quantum
dissipation in such non-equilibrium problems \cite{Next}.

This paper is organized as follows. In section II we
describe our model system (the out of equilibrium spin-fermion
model). Section III presents an overview of
 the analytical results for the non-equilibrium dynamics, along with
the relation to the non-equilibrium x-ray edge problem, while
section IV presents numerical results. In section V we present the
implications for the tunneling rate. Our results imply a breakdown
of the Coulomb gas picture at intermediate times, described in
section VI. In section VII we conclude.

\section{Model}

Our model system consists of a biased two state system (spin)
coupled to two electronic reservoirs $n=L,R$ held at
different chemical potentials. In what follows we assume that the
temperature is zero, and investigate the possibility of  {\it
voltage} activated excitation between the spin states.
The extension to non-zero temperature is straightforward, both analytically
and numerically. The total Hamiltonian is the sum of three terms:

\beq
H=H_S+H_B^{(f)}+H_{SB}^{(f)}.
\label{eq:HTotal}
\eeq
The spin system $H_S$ consists of a two level system (TLS)
(creation operators $d_{\pm}^{\dagger}$) with a bare tunneling
amplitude $\Delta$ and a level splitting $B$. The reservoir term
$H_B^{(f)}$ includes two non-interacting metallic leads $n=L,R$,
where a non-equilibrium state occurs when the leads have different
chemical potentials $\Delta\mu=\mu_L-\mu_R\neq 0$. The system-bath
interaction $H_{SB}^{(f)}$ couples the spin with scattering
processes inside the leads (diagonal coupling), and in between
each lead (nondiagonal coupling), and we choose conventions such
that only one of the spin levels couples to the leads,
\bea
H_S&=&\frac{B}{2}\sigma_z+\frac{\Delta}{2} \sigma_x,
\nonumber\\
H_B^{(f)}&=&\sum_{k,n}\epsilon_k a_{k,n}^{\dagger}a_{k,n},
\nonumber\\
H_{SB}^{(f)}&=&\sum_{k,k',n,n'}V_{k,n;k',n'}a_{k,n}^{\dagger}a_{k',n'}n_d(+).
\label{eq:Hfermion}
\eea
Here $n_d(\pm)= (I\pm\sigma_z)/2$ is the number operator,
with $I$ as the identity operator
\cite{Hamilton}. The operator $a_{k,n}^{\dagger}$ ($a_{k,n}$)
creates (annihilates) an electron with momentum $k$ in the $n$-th
lead. In this paper we focus on the model presented in Ref.
\cite{Ng}, where the momentum dependence of the scattering
potential is neglected. System-bath scattering potentials are then
given by $V_{n,n'}$, where $n,n'=L,R$ are the Fermi sea indices.
Our main conclusions, however, are valid for more general cases.

We assume that
the reservoirs have  the same density of states $\rho(\epsilon)$,
typically modeled using a Lorentzian function
\beq
\rho(\epsilon)=\frac{1}{\pi}\frac{D_L/2}{\left(\frac{D_L}{2}\right)^2+\epsilon^2},
\label{eq:Loren}
\eeq
where $D_L$ is a bandwidth parameter. We typically work in the limit of
wide bands, $D_L\gg\Delta\mu$, therefore to a good approximation
$\rho(\mu_L)\simeq\rho(\mu_R)$.

Note that we ignore the spin degree of freedom of the reservoir electrons in our
discussion. In what follows we refer to the energy difference $B$ as a magnetic field, in
order to distinguish it from the voltage bias $\Delta\mu$.
We also define two auxiliary Hamiltonians $H_{\pm}$ that will be useful
below
\beq
H_{\pm}=\pm \frac{B}{2}+H_{SB}^{(f)}[n_d=\pm]+H_B^{(f)}.
\label{eq:Hpm}
\eeq
Explicitly, $H_{\pm}$ includes the electronic reservoirs and system-bath 
interaction, given that the subsystem  is in the $\pm$ state.
The model (\ref{eq:Hfermion}) contains much of the physics of the Kondo
model \cite{Kondo}, while lacking direct coupling of the
reservoir degrees of freedom to spin-flip processes.
It also contains the spin-resonant-level model of Ref.
\cite{Aditispin} with a particular choice of system-bath couplings.
We discuss the spin-resonant-level model in more detail in Appendix A.

Crucial parameters of the model are the
$L$ and $R$ scattering phase shifts. In equilibrium the
phase shifts are given by \cite{Nozieres,Ng}
\bea \tan \delta_{\pm}= \frac{1}{2}[(\alpha_1+ \alpha_2) \pm
[(\alpha_1-\alpha_2)^2 +4|\nu|^2] ^{1/2}], \label{eq:phaseE} \eea
where $\nu$ and $\alpha_n$ ($n=L,R$) are dimensionless system-bath
coupling strengths
\beq \nu=\pi\rho(\epsilon_F) V_{L,R}; \,\,\,\,\alpha_1=\pi\rho(\epsilon_F) V_{L,L};
\,\,\,\ \alpha_2=\pi\rho(\epsilon_F) V_{R,R}, \eeq
and $\rho(\epsilon_F)$ is the density of states at the Fermi energy of the $L,R$ reservoirs.
Out-of-equilibrium, $\Delta\mu \neq
0$, the phase shifts are {\it complex} numbers given by \cite{Ng},
\bea \tan \delta_L&=&  \alpha_1+i\frac{|\nu|^2}{1-i\alpha_2},
\nonumber\\
\tan \delta_R&=&  \alpha_2-i\frac{|\nu|^2}{1+i\alpha_1}.
\label{eq:phaseC0}
 \eea
Since the reservoirs density of states weakly varies around the
Fermi energy, $D_L\gg \Delta\mu$, the phase shifts are
approximately energy independent, and are all calculated at the
Fermi energy $\epsilon_F $\cite{Ng}. For simplicity, throughout
the paper we typically consider the case of $\alpha\equiv
\alpha_1=\alpha_2\ll \nu$, and take $V_{n,n'}$ to be real. We note
however that our main results, in particular the appearance of
Marcus-type behavior in the non-equilibrium regime at strong
coupling, can be rederived using other variants of this model
system with no limitations on the strength of the diagonal
$V_{n,n}$ interactions,  as well as for the spin-resonant-level
model of Ref. \cite{Aditispin}, see Appendix A.

Under these simplifications, the non-equilibrium phase shifts are given by
\bea \tan \delta_L&=&  \nu^2(i-\alpha),
\nonumber\\
\tan \delta_R&=&  -\nu^2(i+\alpha).
\label{eq:phaseC}
\eea
For $\alpha=0$ the inverse tangent in Eq. (\ref{eq:phaseC}) has a
branch cut, conventionally placed at $(-i\infty,-i]$ and
$[i,i\infty)$. For this special case,
\bea
\delta_L=-\delta_R=\frac{1}{2}\ln\left (\frac{1+\nu^2}{1-\nu^2}\right).
\eea
The weak potential limit therefore corresponds to
$\delta_+=-\delta_-\sim \nu$ and $\delta_L=-\delta_R\sim i\nu^2$,
so $|\delta_{L,R}|\ll |\delta_\pm|$. However, as $\nu \rightarrow
1$, $\delta_{L,R}$ diverges, whereas the equilibrium  phases
$\delta_{\pm}$ are finite.

An important quantity that will be useful below is the sum of
the phase shifts squared, $\tilde \gamma=-(\delta_L^2+\delta_R^2)/\pi^2$.
While for the general model Eq. (\ref{eq:phaseC0}) yields complex numbers,
when the system is symmetric ($\alpha_1=\alpha_2$),
the phase shifts are complex conjugates
and $\tilde \gamma$ is real, although possibly negative.
We discuss the implications of this result in section IV.C .

For the sake of completeness and comparison the
{\it equilibrium} spin-boson model is discussed in Appendix B.
In the nonadiabatic limit, this model yields the
classical Marcus rate at high temperatures when the system-bath
interaction is strong. We analyze the analogous
behavior in the non-equilibrium spin-fermion model
(\ref{eq:Hfermion}) in section V.

\section {nonadiabatic dynamics}

\subsection{Overview}

We are interested in the reduced density matrix $\rho_d(t)$ in the space of
$d$ occupancy. This is defined in terms of time evolution from an initial condition
$\rho_i$ at time $t=0$,
\bea
\rho_d(t)=Tr_{leads}\left[ e^{-iHt}\rho_i e^{iHt}\right].
\eea
If the parameter $\Delta$ in Eq. (\ref{eq:Hfermion}) vanishes, the
problem is just electrons in a time-independent potential, and a
closed form analytical solution exists. If $\Delta \neq0$, $\rho_d(t)$
may be expressed as an expansion in $\Delta$. Evaluation of any
term in the expansion entails solving a problem of electrons in a
{\it time-dependent} field. In equilibrium, an essentially exact
closed-form solution exists, and the main problem is to re-sum the
series in $\Delta$ \cite{Chen,Chang-Chak}. For non-equilibrium problems an analytical
expression is not known. In this paper we present a detailed
numerical evaluation of some low order terms in the expansion for
$\Delta$. The essential features are revealed by the Golden Rule
decay rate, obtained by assuming that (i) $n_d(+)=0$ at time
$t=0$, and (ii) $\rho_i$ is the density matrix corresponding to
the ground state of $H$ with $n_d(+)=0$, and (iii) that an expansion
to ${\cal O}(\Delta^2)$ suffices.
This level of description is equivalent to the "non-interaction
blip approximation" (NIBA) in the standard spin-boson model
\cite{WeissBook,Leggett,Aslangul} and yields the
(nonadiabatic) Fermi Golden Rule for the forward ($+$) and
backward ($-$) transition rates between the spin levels as
\cite{Mahan,Lax,Kubo,Golosov}
\bea \Gamma_f^{\pm}&=&\left(\frac{\Delta}{2}\right)^2 2 \Re
\int_{0}^{\infty} e^{\pm iBt}C_f(t) dt;
\nonumber\\
C_{f}&=& e^{iE_-t} \langle T e^{-i\int_0^{t} d \tau H_{SB}^{(f)}(\tau, n_d=+)} \rangle
\nonumber\\
&=&\langle e^{-iH_+ t} e^{iH_- t}\rangle \equiv e^{-\Phi_f(t)},
\label{eq:gaf}
\eea
%
%
%
where $T$ denotes time ordering,
$H_{SB}^{(f)}(t)=e^{iH_B^{(f)}t}H_{SB}^{(f)}e^{-iH_B^{(f)}t}$,
and  $E_-$ is the ground state energy of the two uncoupled reservoirs.
The Hamiltonians $H_{\pm}$ are defined in Eq. (\ref{eq:Hpm}),
$\Re$ refers to the real part of the integral, and the trace is
performed over the electronic degrees of freedom. 
For convenience, the term including the energy bias $B$ is taken outside of the trace.

The object of our calculation
is therefore the correlation function $C_f(t)$, which should be
evaluated for non-equilibrium conditions covering time scales from
$Dt\sim 1$ up to $\Delta\mu t \gg 1$.
$D$, an energy of the order of the Fermi seas bandwidth, and the
potential drop $\Delta\mu$, specify two inverse time scales in the
problem, where we typically work in the limit of  $\Delta\mu\ll D$.
Note that, unlike the equilibrium case,
there is no exact analytical approach to
calculate $C_f(t)$ valid for all time scales.
Approximate analytical approaches and exact numerics may
be performed, as discussed below.

\subsection{Short and long-time asymptotics;
Non-equilibrium x-ray edge problem}

The correlation function $C_f(t)$  [Eq. (\ref{eq:gaf})] is a
crucial element in the theory of the x-ray edge problem,
an effect originating from
 the many body response of a Fermi system to the fast switching of a
scattering potential, e.g. the creation of a core hole \cite{Nozieres,Ohtaka}.
The x-ray edge Hamiltonian is a simplified version of the spin-fermion
model, Eq. (\ref{eq:Hfermion}), with a static subsystem that is
either empty or populated,
\bea H_0&=& H_B^{(f)} +\epsilon_dd^{\dagger}d,
\nonumber\\
H_{SB}&=&\sum_{k,n=L,R}V_{k,n;k',n'}a_{k,n}^{\dagger}a_{k',n'}d^{\dagger}d,
\nonumber\\
H_{edge}&=&H_0+H_{SB}.
\label{eq:edge}
 \eea
Here $d^{\dagger}$, $d$ are creation and destruction operators of
the core electron, and $a_{k,n}^{\dagger}$ ($a_{k,n}$) creates
(destroys) an electron in the $n$-th lead with momentum $k$. The
{\it single} band x-ray singularity problem was originally solved
exactly in the asymptotic limit by Nozieres and De Dominicis (ND)
\cite{Nozieres}.
%
%
%

In the last ten years there has been a growing interest in
understanding the x-ray edge effect in the mesoscopic regime
\cite{LevitovM,Baranger} and for {\it non-equilibrium} systems
\cite{Ng,Combescot,Braunecker0,Braunecker1,Muzy2,Braunecker2,Levitov},
where the core hole couples to more than one Fermi sea at
different chemical potentials. Standard equilibrium techniques,
e.g. bosonization \cite{Schotte,Giamarchi,Hakim}, cannot be simply
generalized to handle these non-equilibrium systems (see Appendix
C). The first to address the non-equilibrium problem was Ng, who
generalized the Nozieres-De Dominicis solution to include more
than one Fermi sea with different chemical potentials \cite{Ng}.
Ng demonstrated that the edge singularity could be described by
generalized phase shifts which are real for equilibrium systems
and complex when the system is driven out-of-equilibrium.
Physically, complex phase shifts reflect the finite lifetime of a
non-equilibrium system. More recently, Muzykantskii {\it et al.}
\cite{Braunecker1,Muzy2} formally solved the out-of-equilibrium
problem using the Riemann-Hilbert approach. The result, given in
terms of the scattering matrix, was later generalized to include
finite temperature effects \cite{Braunecker2}. An exact formal
determinant solution was presented in Ref. \cite{Levitov} for the
study of tunneling in a non-equilibrium electron gas.

A formal solution for $C_f(t)$ is obtained from
the linked cluster theorem (valid also for non-equilibrium problems)
\cite{Nozieres,Ng},
\bea
\Phi_f(t)=-\int_0^{1} d\lambda Tr  \underline {\underline V}(t) \underline {\underline {G}}^{\lambda}(t,t),
\eea
with $\underline{\underline G}^{\lambda}(t,t')$ the matrix Green function in the space of the
leads for Eq. (\ref{eq:HTotal}), but with  $H_{SB}\rightarrow \lambda H_{SB}$.
For this model  $\underline {\underline G}$ solves the Dyson equation
\bea
\underline {\underline G}^{\lambda}(t_1,t_2)
=\underline {\underline g}(t_1,t_2)+ \int d \tau \underline {\underline g}(t_1,\tau)
\underline{ \underline V}^{\lambda}(\tau)
\underline {\underline G}^{\lambda}(\tau,t_2),
\label{eq:G}
\eea
with  $\underline{ \underline V}^{\lambda}(\tau)=V_{n,n'}n_d(+)(\tau)$
and the unperturbed Green's functions
\bea
g_{n,n'}(t_1,t_2)=\frac{\pi \rho}{t_1-t_2}e^{i(\mu_nt_1-\mu_{n'}t_2)}.
\eea
$\rho$ is the reservoir density of states,
taken to be the same for the $L$ and $R$ leads.
%
%

In equilibrium, ND showed \cite{Nozieres} that this equation can be solved exactly
and the coupling constant integral performed, leading to
\bea
C_f(t)\sim (Dt)^{-\beta},
\label{eq:Ceq}
\eea
with $D$ an energy of the order of the Fermi sea bandwidth,
$\beta=(\delta_+^2+\delta_-^2)/\pi^2$, and $\delta_\pm$ defined in
Eq. (\ref{eq:phaseE}). For the $\alpha=0$ model studied
explicitly
\bea \beta=2\left(\frac{\rm {atan}(\nu)}{\pi}\right)^2.
\label{eq:betadef} \eea
Eq. (\ref{eq:Ceq}) also holds for the non-equilibrium problem  at
times $\Delta \mu t \ll1$.

At long times, $\Delta\mu t \gg 1$, the equation was solved by Ng \cite{Ng};
see also \cite{Braunecker1}. The coupling
constant integral may similarly be performed, leading to
\bea C_f(t) \sim e^{-\Gamma \Delta\mu t} (\Delta\mu t)^{\gamma},
\label{eq:CNeq} \eea
with $\Gamma=|\delta_L''-\delta_R''|/2\pi$, $\gamma=-(\delta_L^2+\delta_R^2)/\pi^2$ and
$\delta_{L,R}$ given by Eq. (\ref{eq:phaseC}).
Here $\delta_n''$ ($n=L,R$) refers to the imaginary part of the
phase shift.
For the model studied numerically ($\alpha=0$) we have
\beq \Gamma
= \frac{1}{2\pi} \ln
\left[\frac{1+\nu^2}{1-\nu^2}\right],
\label{eq:gamma} \eeq
and
\beq
\gamma =
\frac{1}{2\pi^2} \ln^2  \left[\frac{1+\nu^2}{1-\nu^2} \right].
\label{eq:gamacof}
\eeq
%


\subsection{Intermediate time}

Although the long time and short time behavior is known
essentially exactly, a transparent non-perturbative analytical
expression for $C_f(t)$ that encompasses all time scales
$\Delta\mu t$ and coupling strengths $\rho V$ has not been
developed. Indeed, in this work we argue that at strong coupling
($\nu \rightarrow 1$) a different functional form dominates at
{\it intermediate times} $\Delta\mu t \sim 1-10$ where a prominent
{\it Gaussian} decay emerges, $C_f(t) \sim e^{-\kappa(\Delta\mu
t)^2} (Dt)^{-\beta}$. We first offer a perturbative  calculation
which suggests this result, and then present exact numerical
simulations which prove this behavior. The dominance of the
Gaussian behavior at intermediate times translates into a
Marcus-type rate in frequency domain, with {\it bias voltage}
activation (see discussion in section V),  instead of {\it
temperature} activation, as in the classical Marcus rate (see
Appendix B).


The correlation function $C_f(t)$ can be evaluated using the
cumulant expansion \cite{Mahan}. Note that unlike the bosonic
case, {\it all} cumulants contribute,
\bea &&C_f(t)=\exp \sum_{n=1}^{\infty} K_n(t);
\nonumber\\
&& K_n(t)=
\nonumber\\
&&\frac{(-i)^n}{n!} \int_0^t dt_1 \int_0^t dt_2... \int_0 ^t dt_n \langle T F(t_1)F(t_2)...F(t_n)\rangle _c,
\label{eq:expansion}
\nonumber\\
\eea
where $T$ denotes time ordering,
$F=\sum_{k,k',n,n'}V_{n,n'}a_{k,n}^{\dagger}a_{k',n'}$, and
$\langle... \rangle_c$ denotes a cumulant average.
The first cumulant yields an energy shift, while the second term is
given explicitly by ($Dt>1$, $\alpha=0$)
\bea K_2(t)&=& -\frac{1}{2} \int_{0}^{t} dt_1\int_0^t dt_2\langle
T F(t_1) F(t_2)\rangle_c
\nonumber\\
&=& - \frac{2\nu^2}{\pi^2}\ln(Dt)
\nonumber\\
&-&2\frac{\nu^2}{\pi^2} \Delta\mu t \bigg [{\rm Si}(\Delta\mu t) -\frac{1-\cos(\Delta\mu t)}{\Delta \mu t}\bigg ]
\nonumber\\
&+&2 \frac{\nu^2}{\pi^2}[\gamma_e+\ln(\Delta\mu t)- {\rm Ci}(\Delta\mu t)].
\label{eq:K2}
\eea
For details see Appendix D.
The sine and cosine integrals are
defined as ${\rm Si}(x)=\int_0^{x} \frac{\sin(t)}{t}dt$, ${\rm
Ci}(x)= \gamma_e+\ln(x) +\int_0^{x} \frac{\cos(t)-1}{t}dt$, and
$\gamma_e=0.5772$ is the Euler-Mascheroni constant.

This expression reproduces the weak coupling limits of the analytical
results Eqs. (\ref{eq:Ceq}) and (\ref{eq:CNeq}) at short and long
times respectively, and provides an interpolation between the two
times. In particular, the second line describes how the long-time
dissipation $\Gamma\Delta\mu t$ term is "turned on" as $\Delta\mu
t$ increases from a small value to values much greater than unity.
The first and last terms describe how the equilibrium
orthogonality is turned off as $\Delta\mu t$ increases:  $[
\gamma_e+\ln(\Delta\mu t) -{\rm Ci}(\Delta\mu t) - \ln(Dt)]$ is a
function which interpolates between $\ln (t)$ for
$1/D<t<1/\Delta\mu$, and a constant at $\Delta\mu t\gg 1$.
Note that in this model the leading logarithmic term at long times
is $\sim \nu^4$, consistent with the cancellation of logarithmic
terms in the long time limit of Eq. (\ref{eq:K2}), i.e. with the
absence of a term which "turns on" the non-equilibrium power law.
This cancellation does not necessarily occur at order $\nu^2$ in
other models, e.g. the spin-resonant-level model, see Appendix A.
%
%
%

%
We can clearly distinguish between three regimes in Eq.
(\ref{eq:K2}):
\bea
C_{f}(t) \sim
\begin{cases}
t^{-2\nu^2/\pi^2} & \Delta\mu t \ll 1  \\
e^{-(\nu \Delta \mu t)^2/2\pi^2} \times t^{-2\nu^2/\pi^2} & \Delta\mu t \sim 1  \\
e^{-\nu^2  \Delta\mu t /\pi } & \Delta\mu t \gg 1. \\
\end{cases}
\label{eq:cases}
\eea
While the first (equilibrium) limit and the third regime are well
established in the literature \cite{Ng,
Braunecker1,Aditispin,Muzy2}, the intermediate domain, leading to
an interesting new dynamic has not been discussed. In the strong
coupling limit the Gaussian behavior may have a dominant effect on
the relaxation, as discussed below. We would like therefore to
phenomenologically extend the second cumulant expression, Eq.
(\ref{eq:K2}), to larger phase shifts (strong coupling).


Perturbative expressions analogous to Eq. (\ref{eq:K2}) motivated Mitra and Millis
\cite{Aditisemi} to propose an interpolation function
constructed by replacing the factors of $\nu^2$ in the expression above by the exact phase shifts.
For the model considered here their procedure leads to
\bea
&&\Phi_{f}(t)=
\nonumber\\
&&+\frac{|\delta_L''- \delta_R''|}{\pi^2} (\Delta\mu t)
 \left[\rm{Si}(\Delta\mu t) -\frac{1-\cos(\Delta\mu t)}{\Delta\mu t} \right]
\nonumber\\
&&+
\frac{(\delta_+^2+\delta_-^2)}{\pi^2}
\left[\ln(1+iDt) -\gamma_e+{\rm Ci}(\Delta\mu t) -\ln(\Delta\mu t) \right]
\nonumber\\
&&+ \frac{ (\delta_L^2 +\delta_R^2)}{\pi^2} \left[\gamma_e-{\rm Ci}(\Delta \mu t)+\ln(\Delta\mu t) \right].
\label{eq:Rfer}
\eea
%
Note that our approximation for the scattering potentials,
$\alpha_1=\alpha_2=0$, implies that there is no Fumi energy shift.
 At the short time limit, $\Delta\mu t \ll1$, the factor
$[\ln(\Delta\mu t)+\gamma_e-{\rm Ci(\Delta\mu t)}]$ dies out,
leading to the correct equilibrium behavior (\ref{eq:Ceq}).
%
%
In contrast, at long times the cosine integral diminishes, which
implies that the dynamics is ruled by an exponential decay with a
rate constant $\Delta \mu \Gamma$, [Eq. (\ref{eq:gamma})],
modified by a power law term  $t^{\gamma}$,
Eq. (\ref{eq:gamacof}).
%
%
%
%

Our numerical results, to be presented below, show that at weak to
moderate coupling, $\nu <0.5$, the correlation function and the
resulting transition rates are well described by expression
(\ref{eq:Rfer}). However, Eq. (\ref{eq:Rfer}) is found to be a
poor approximation at strong coupling.
Instead, at intermediate times $\Delta\mu t \sim 1$ we return to
Eq. (\ref{eq:K2}) and replace
the weak coupling phase shift by the {\it equilibrium} strong
coupling phase shift, $\nu\rightarrow \rm {atan} (\nu)$. The
physical picture is that on these time scales the phase shifts are
essentially still the {\it equilibrium} ones. Only at longer times $\Delta\mu t \gg 1$ the
non-equilibrium dynamics is reflected in the complex phase shifts
(\ref{eq:phaseC}). This conjecture yields
\beq
\Phi_f(\Delta\mu t \sim1)= \Phi_{eq}(t) + \Phi_{neq}(t)+iE_st,
\label{eq:3t}
\eeq
where the equilibrium function is the same as in the zero bias
case \cite{Nozieres},
\bea
\Phi_{eq}(t)&=& \beta\ln(1+iDt);
\nonumber\\
\beta&=& \frac{(\delta_+^2+\delta_-^2)}{\pi^2}=
2 \frac{{\rm atan}^2 (\nu)  }{\pi^2},
\label{eq:phieq}
\eea
while the  non-equilibrium term provides a quadratic time decay
\bea
\Phi_{neq}(t)&=&  \kappa (\Delta\mu t)^2;
\nonumber\\
\kappa&=&\frac{(\delta_+^2+\delta_-^2)}{4\pi^2}
=\frac{\rm{atan}^2(\nu)}{2 \pi^2} .
\label{eq:phineqkappa}
\eea
Notice that the prefactor $\kappa$ depends only on the scattering
potential $V_{n,n'}$.
The last element in Eq. (\ref{eq:3t}) is the
energy shift $E_s$. We assume that it is given by the equilibrium  limit
of the Fumi's theorem,
%
\bea
E_s=\frac {D} {\pi} [\delta_-+\delta_+].
\label{eq:Esint}
\eea
For $\alpha=0$ the energy shift is zero.

Similarly to expressions (\ref{eq:3t})-(\ref{eq:Esint}), Eq.
(\ref{eq:Rfer}) gives the first correction to the equilibrium
result which is proportional to $(\Delta\mu t)^2$, but in contrast
to these equations, the coefficient involves
 the {\it non-equilibrium} exponent,
and in fact does not provide a wide regime of $t^2$ behavior.
%
Our numerical simulations, presented below, support expressions
(\ref{eq:3t})-(\ref{eq:phineqkappa}) in the broad window
$\Delta\mu t \sim 1-10$. We have not been successful in
constructing a general analytical expression, valid on all time
scales and coupling strengths. It is possible that consideration
of the fourth cumulant may yield some insight here. 

\section{Numerics \label{Numerics}}

\subsection{Methods}

The fermionic correlation function $C_f(t)$ can be directly
calculated by expressing the zero temperature many body average as
a determinant of the single particle correlation functions
\cite{Mahan,Temp}
\bea C_f(t)&=&\langle e^{-iH_+t} e^{iH_-t} \rangle
\nonumber\\
&=& \det \left[\phi_{k,n; k',n'}(t)  \right]_{k<k_f^n;
k'<k_f^{n'}};
\nonumber\\
\phi_{k,n;k',n'}(t)&=&\langle k,n| e^{-ih_+t}e^{ih_-t} |k',n'
\rangle. \label{eq:num} \eea
Here $H_{\pm}=\sum h_{\pm}$, where $h_{\pm}$ are the single
particle Hamiltonians for the individual conduction electrons.
$|k,n\rangle$ are the single particle eigenstates of $H_B^{(f)}$,
and the determinant is evaluated over the occupied states.
$k_f^{n}$ is the Fermi energy of the $n$-th reservoir. In our
numerical calculations we have used a Lorentzian density of
states, with  tails that are long enough to eliminate artificial
reflections from the boundaries. The Lorentzian function is
centered around the equilibrium Fermi energy with a full width at
half maximum $D_L$, Eq. (\ref{eq:Loren}). This quantity sets
energy and time scales in our simulations. We have typically used
$D_L=4.5$ for the two reservoirs, $\Delta\mu/D_L<0.1$ and
$\nu=0.1-1$. We also take the diagonal coupling to be zero
($\alpha=0$) in all of our simulations, unless otherwise stated.
For these parameters, we have found that for  short-time
evolution ($\Delta\mu t<15$), even for strong coupling, it is
satisfactory to model the fermionic reservoirs using $\sim$400
states per bath, where bias is applied by depopulating one of the
reservoirs with respect to the other.

We can also employ the renormalization group (RG) method,
originally developed by Wilson for the calculation of the
thermodynamic properties of the Kondo problem \cite{Wilson}, for
the numerical solution of the {\it non-equilibrium} x-ray edge
problem. In equilibrium, Oliveira {\it et al.}
\cite{Oliveira1,Oliveira2} have used the RG technique to calculate
the x-ray absorption spectrum. This study can be generalized to
include two reservoirs with different chemical potentials by
following a three-step procedure: (i) define the conduction bands
on a logarithmic scale. (ii) convert the (isolated) reservoir
Hamiltonians into semi-infinite tight binding chains, as is
done in Refs. \cite{Oliveira1,Oliveira2}. In this
representation the impurity couples the chains' first levels.
(iii) build the Hamiltonians $H_{\pm}$ in the new basis, first
including the occupied levels of the $L$ and $R$ reservoirs, then
adding the empty levels. The determinant (\ref{eq:num}) is
performed over occupied levels only.

\begin{figure}
{\hbox{\epsfxsize=80mm \epsffile{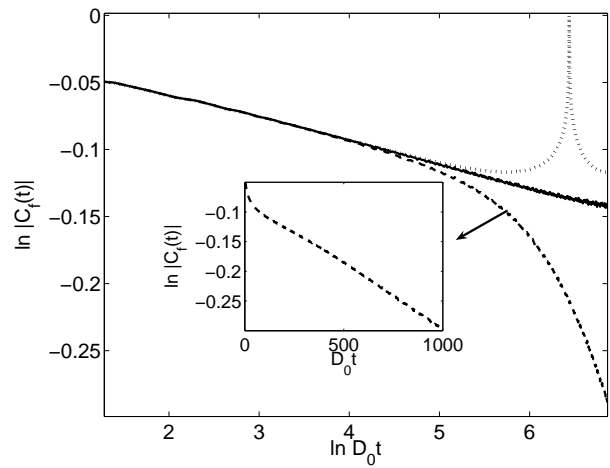}}} \caption{Decay of
$C_f(t)$ as computed by both the RG approach and standard linear
discretization for $\nu=0.1$.  Constant discretization with 200
states per band, $\Delta\mu=0$ (dotted), showing an artificial
rise of the correlation function due to discretization errors;
 logarithmic discretization (RG) with 100 states per band, $\Lambda=1.1$,
$\Delta\mu=0$ (full); logarithmic discretization (RG) with 100
states per band, $\Lambda=1.1$, $\Delta\mu=0.007$ (dashed).
$\Lambda$ is a logarithmic scale parameter, where for each
conduction energy there are states with energies
$\epsilon/\Lambda^m$, $m=1,2...$. Inset: The finite bias case
exhibits an exponential decay at long times. The slope agrees with
the theoretical value, Eq. (\ref{eq:gamma}). The Fermi sea
bandwidth is $2D_0$ in all plots with $D_0=1$. }
 \label{FigRG}
\end{figure}

In equilibrium, the RG technique is highly advantageous over
constant/Lorentzian discretization methods, as it converges
rapidly to the continuum limit
even for gross discretization. 
For small voltage differences ($\Delta\mu/D <10^{-2}$) this method
nicely reveals the crossover of $C_f(t)$ from equilibrium to
non-equilibrium behavior with increasing bias.
In contrast, for large bias the Lorentzian discretization is more
convenient, since energies far from the Fermi energy are not well
represented within the RG technique.
We present a numerical example in Fig. \ref{FigRG}, demonstrating
the strength of the RG approach over standard linear
discretization for systems in equilibrium. The RG technique
provides stable dynamics for long times (full line), where
constant discretization fails (dotted line), yielding an
artificial rise of the correlation function due to discretization
errors. The theoretical value of $\beta=2 {\rm atan}^2(\nu)/\pi^2
=0.0020$ for $\nu=0.1$, nicely agrees with the numerical slope of
$0.0019$.  Deviations are due to the sharp energy cutoff used at
$D_0$=1, with the conduction band energies extending from $-D_0$
to $D_0$. We also present the results of an RG calculation  with a
very small voltage drop (dashed line), where linear discretization
would require a very fine grid.

In this work we typically focus on systems far from equilibrium,
$\Delta\mu/D\sim 0.1$. Since the RG method samples  the Fermi sea
states predominantly near the Fermi energy, while high energy states
are under-represented, we find the Lorentzian discretization to be more convenient.

\subsection{Results: $C_f(t)$}

\begin{figure}
{\hbox{\epsfxsize=80mm \epsffile{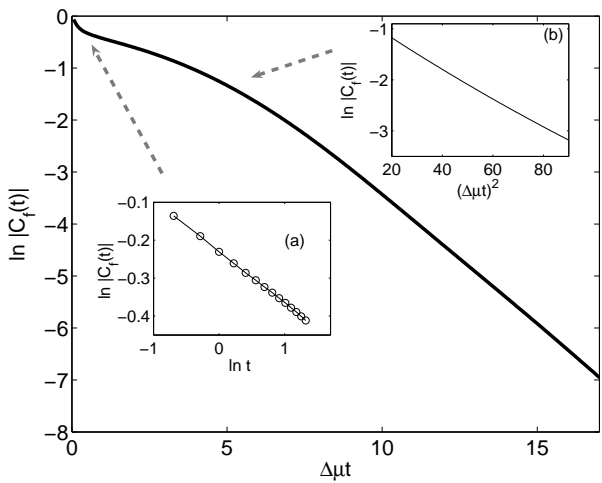}}} \caption{The correlation
function $C_f(t)$ for $\Delta\mu$=0.24, $\nu$=0.95, manifesting a
power law decay at short times $\Delta\mu t<1$ (a) (notice the log-log scale),
a Gaussian decay at intermediate times $\Delta\mu t\sim 1-10$ (b), and an
exponential decay at long times $\Delta\mu t\gg1$.
}
 \label{Fig1}
\end{figure}

Representative results are
displayed in figure \ref{Fig1}. The main plot presents the
logarithm of the correlation function $|C_f(t)|$ at strong
coupling $\nu=0.95$ for an applied voltage $\Delta\mu=0.24$.
%
Three different regimes are clearly identified: a power law decay
at short times $\Delta\mu t<1$, see lower left inset (a), an
exponential decay at long times $\Delta\mu t \gg 1$, and
remarkably,  an intermediate regime $1\lesssim\Delta\mu t\lesssim10$ of
approximately Gaussian behavior [upper right inset (b)]. The short
and long time behaviors are consistent with the theoretical
results. The intermediate time quasi-Gaussian regime is a new
finding with important consequences.
We  analyze the short time
dynamics $\Delta\mu t<1$, enlarged in Fig. \ref{Fig1}(a),  by
fitting the data to the analytic expression  $\ln C_f(t) \sim
-\beta\ln(Dt)$. This provides an effective bandwidth $D=6$ and a
decay constant $\beta=0.13$ consistent, within numerical errors,
with the theoretically expected $\beta=2[{\rm atan }(\nu)/\pi]^2
\approx 0.12$.
We can also fit the intermediate time
behavior, shown in Fig. \ref{Fig1}(b), by a Gaussian function $\ln
C_f(t) \sim -\kappa \Delta\mu^2 t^2$ which yields the prefactor
$\kappa$=0.03.

\begin{figure}
{\hbox{\epsfxsize=80mm \epsffile{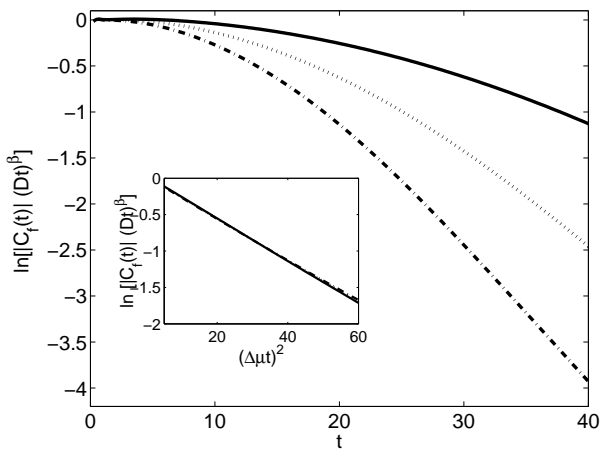}}} \caption{Evidence for the
Gaussian decay at intermediate times $\Delta\mu t\sim$ 1-10 and
strong coupling $\nu=0.95$, $\Delta\mu=0.16$ (full),
$\Delta\mu=0.24$ (dotted) and $\Delta\mu=0.32$ (dashed-dotted).
The inset, which includes the three lines one on top of the other,
reveals that $C_f(t) t^{\beta}\propto e^{-\kappa (\Delta\mu
t)^2}$, $\beta$=0.13, with $\kappa\sim 0.03$.
} \label{Fig4}
\end{figure}

Figure \ref{Fig4} presents a more detailed examination of the
Gaussian behavior, showing that at both short and intermediate
times, the data can be well described by the approximate function
\bea
y_G(t)=(Dt)^{-\beta}e^{-\kappa (\Delta\mu t)^2},
\label{eq:yG}
\eea
with $\beta$ the theoretically predicted short time (equilibrium)
exponent $(\delta_+^2+\delta_-^2)/\pi^2$. The inset proves that
the data follows the same linear trend when plotted as a function
of $(\Delta\mu t)^2$, with a slope of $\kappa \sim 0.03$. This
value nicely agrees with the constant predicted by Eq.
(\ref{eq:phineqkappa}), $\kappa={\rm atan}(\nu)^2/2\pi^2$=0.029
($\nu=0.95$).

\begin{figure}
{\hbox{\epsfxsize=80mm \epsffile{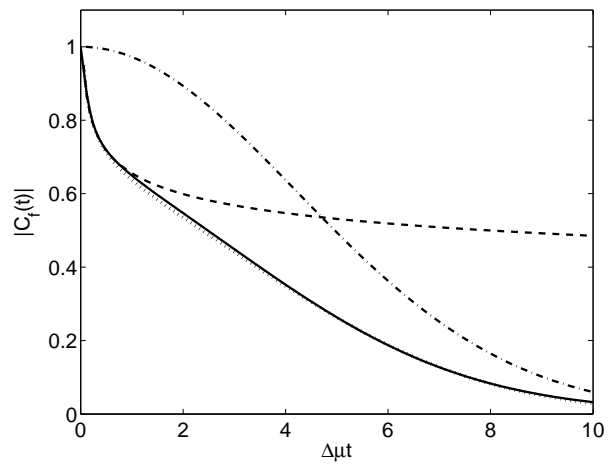}}} \caption{Time dependence
of the correlation function $|C_f(t)|$ for $\Delta\mu=0.24$,
$\nu=0.95$. The fitting function $y_G=e^{-\kappa(\Delta \mu t)^2}
(Dt)^{-\beta}$ (dotted) with $\kappa=0.03$, $D=6$ and $\beta=0.13$
compared to exact numerical solution (full). The Gaussian
(dashed-dotted) and the power law (dashed) parts of $y_G$ are also
displayed separately.
}
 \label{Fig5}
\end{figure}

Fig. \ref{Fig5} provides more insight by
deconstructing the observed time decay of $C_f(t)$ into the
equilibrium power law and non-equilibrium Gaussian components.
 Another important observation deduced from Figs.
\ref{Fig1}, \ref{Fig4} and \ref{Fig5} is that the correlation
function decays to $\sim 0.1$ its initial value by the time the
exponential decay begins to dominate. This implies that the
Gaussian behavior governs the rate constant at strong enough
coupling, leading to a voltage activated regime analogous to the
high-temperature semiclassical polaron transport regime. We call
this "fermionic Marcus" behavior.

\begin{figure}
{\hbox{\epsfxsize=80mm \epsffile{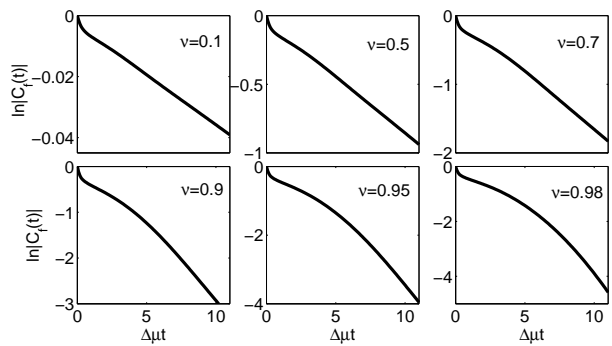}}} \caption{ The correlation
function $C_f(t)$ for different coupling  strengths $\nu$,
manifesting the increasing dominance of intermediate time
 Gaussian behavior at strong coupling. $\Delta\mu=0.24$ in all plots.}
 \label{Fig2}
\end{figure}

Fig. \ref{Fig2} presents the evolution of the correlation
function $C_f(t)$ as coupling strength $\nu$  is varied from
weak to strong. All other
parameters are the same as in Fig. \ref{Fig1}. For all coupling
strengths the short time logarithmic and the approximate long time
exponential behavior are observed. However, as the coupling
strength is increased, increasingly wide intermediate regime is
observed. We have verified, by an analysis similar to that shown
in the lower left inset of Fig. \ref{Fig1}, that the short time
behavior is always a power law with the theoretically predicted
exponent $\beta=2 [{\rm atan}(\nu)/\pi]^2$.
Also note that
 while at weak coupling ($\nu<0.5$) the correlation
function weakly decays before the turnover to an exponential decay
takes place, for very strong coupling, $\nu\gtrsim0.9$, the
dynamics is critically controlled by the Gaussian form, as the
correlation function has decayed to zero before the exponential
decay takes place. This implies that the resulting decay rate [Eq.
(\ref{eq:gaf})] essentially shows different characteristics in
these two regimes.


\begin{figure}
{\hbox{\epsfxsize=80mm \epsffile{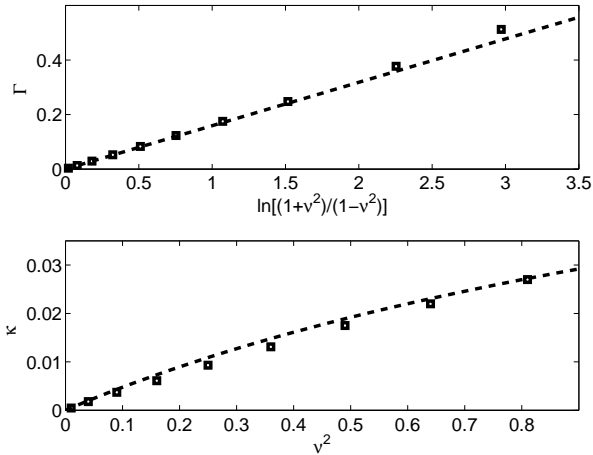}}} \caption{Testing the
validity of Eqs. (\ref{eq:gamma}) and (\ref{eq:phineqkappa}) for
describing the long time and intermediate time behavior,
respectively. Top: The relaxation rate $\Gamma$. Numerical
results, calculated from the slope of $\ln[C_f(t) t^{\beta}]$ vs.
$\Delta\mu t$ at long times (squares); analytical results using
Eq. (\ref{eq:gamma}) (dashed line).
Bottom: The coefficient $\kappa$. The numerical slope of
$\ln[C_f(t) t^{\beta}]$ vs. $\Delta\mu^2t^2$ at intermediate
times (squares);
analytical results using Eq. (\ref{eq:phineqkappa}) (dashed line).
These data were computed with $\Delta\mu$=0.24, and $\beta$ was
extracted from the short time dynamics for
each value of $\nu$. }
\label{Fig3}
\end{figure}

We now systematically explore the Gaussian decay at intermediate
times, and the exponential decay rate at long times, and compare
the numerical coefficients $\Gamma$ and $\kappa$ with the
theoretical values, Eqs. (\ref{eq:gamma}) and
(\ref{eq:phineqkappa}), respectively. This is done by calculating
the correlation function $C_f(t)$ for coupling strengths
$\nu=0.1-0.95$ (see Fig. \ref{Fig2}), then extracting both the
quadratic intermediate slope $\kappa\Delta\mu^2$ and the long time
exponential slope $\Gamma\Delta\mu$. Fig. \ref{Fig3} presents
these coefficients showing excellent agreement with the values
predicted from the phenomenological ansatz, Eqs.
(\ref{eq:3t})-(\ref{eq:phineqkappa}).
%

\begin{figure}
{\hbox{\epsfxsize=80mm \epsffile{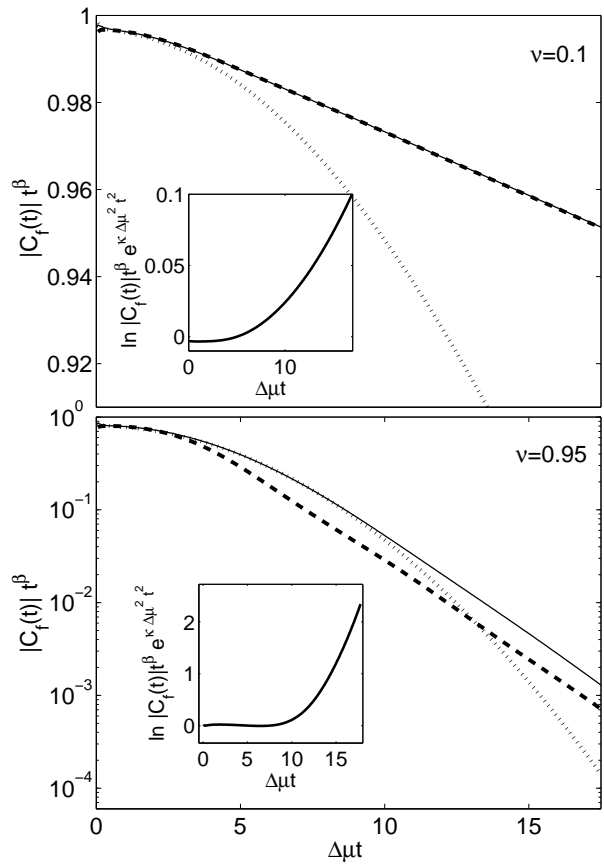}}} \caption{The
turnover from a Gaussian decay to an exponential relaxation.
Top: Comparison between the numerical
correlation function $C_f(t)$ (full) and the fitting functions
$y_G$ (dotted) and $y_E$ (dashed) defined in the text, $\nu=0.1$,
$\Delta\mu$=0.24.
%
Inset: Exposing the turnover by
taking away the short time and intermediate time terms $t^{-\beta}e^{-\kappa
\Delta\mu^2 t^2}$ with $\beta$=0.002, $\kappa=5 \times 10^{-4}$.
Bottom: Same with $\nu=0.95$, leading to $\beta=0.13$, $\kappa=0.03$.
While $y_G$ explicitly includes the Gaussian decay, correct to
$\Delta\mu t\sim 10$, the function $y_E$ captures the correct
slope at longer times.
}
 \label{Fig6}
\end{figure}

Next, in Fig. \ref{Fig6} we examine the crossover to the analytic
long time behavior, Eq. (\ref{eq:Rfer}). We compare the numerical
correlation function with two functions: the approximate fitting
function $y_G$ defined above, and the {\it long time} perturbation
theory result of Ref. \cite{Aditispin}, given by exponentiating
Eq. (\ref{eq:Rfer}). We refer to this second function as $y_E$. We
see that $y_E$ describes the data well at
long times,  but that as the coupling strength is increased, the
range over which the Gaussian description applies increases. This
feature can be qualitatively described by the approximate
crossover function
\bea
C_f^{(app)}(t) \sim \exp\left\{-\left[(\Gamma\Delta\mu t)^2+\frac{\Gamma^4}{4\kappa^2} \right]^{1/2}+\frac{\Gamma^2}{2\kappa}\right\},
\label{eq:At}
\nonumber\\
\eea
which captures the crossover from a Gaussian dynamics to an exponential decay.
An increase of $\nu$ leads to a strong enhancement of $\Gamma$,  while $\kappa$
reaches saturation, resulting in a counterintuitive lengthening of the range of the
intermediate Gaussian dynamics with increased $\Gamma$.

In summary, we have shown that the crossover between  equilibrium
($\Delta\mu t \ll 1$) and non-equilibrium ($\Delta\mu t \gg 1$)
behavior is described by a regime of Gaussian relaxation
negligible for weak coupling, but for strong coupling extending over
the wide range $1 \lesssim \Delta\mu \lesssim10$, with parameters
determined by the equilibrium exponents. In Section V we
examine the consequences for spin relaxation.

\subsection{Orthogonality and anti-orthogonality}

We focus next on the power law contribution to Eq. (\ref{eq:CNeq}).
Unlike the standard equilibrium case, where the
system always experiences dephasing, $\gamma<0$ \cite{Anderson},
in our model the power law term in Eq. (\ref{eq:CNeq}) acquires a positive
exponent $\gamma>0$, {\it enhancing} the correlation function,
see Eq. (\ref{eq:gamacof}).
We refer to this situation as an "anti-orthogonality" effect. For a
general system-bath coupling model, Eq. (\ref{eq:Rfer}) reveals that
\bea
C_f(\Delta\mu t \gg 1) \sim e^{-\Gamma \Delta\mu t}
t^{\tilde \gamma}; \,\,\, \tilde \gamma=-(\delta_L^2+\delta_R^2)/\pi^2,
\label{eq:longgg}
\eea
with complex, non-equilibrium, phase shifts given by Eq.
(\ref{eq:phaseC0}) \cite{Ng}.
It is clear that in the special
limit of zero diagonal interactions ($\alpha=0$), the phase shifts
$\delta_{L,R}$ are purely imaginary and
$(\delta_L^2+\delta_R^2)<0$ for all values of $\nu$, leading to
$\tilde \gamma>0$ . In contrast, for large diagonal coupling we
typically find that $\tilde \gamma<0$, which is the standard
orthogonality behavior. {\it The anti-orthogonality effect is therefore
a footprint of a non-equilibrium situation.}

\begin{figure}
{\hbox{\epsfxsize=80mm \epsffile{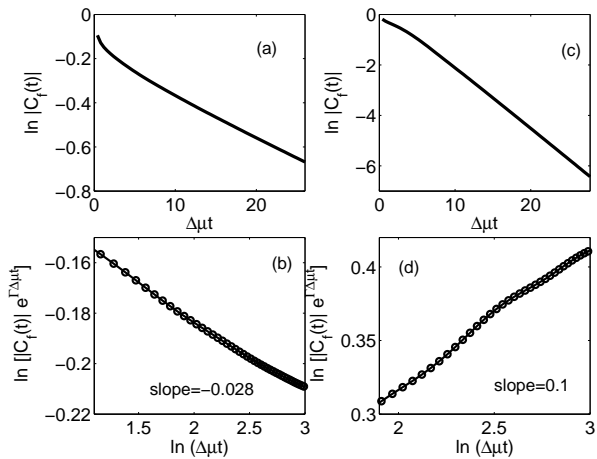}}} \caption{
Orthogonality and anti-orthogonality effects in the non-equilibrium
system $\Delta\mu =0.4$. (a)-(b) $\nu$=0.3, $\alpha$=0.5,
manifesting the standard orthogonality effect ($\tilde \gamma<0$).
(c)-(d) $\nu$=0.8, $\alpha$=0, revealing the anti-orthogonality
effect ($\tilde \gamma>0$). } \label{Fig7}
\end{figure}

We next turn to a numerically exact exploration of the anti-orthogonality
effect.
Fig. \ref{Fig7} shows the correlation
function $C_f(t)$ at long times $\Delta\mu t\gg 1$ when expression
(\ref{eq:Rfer}) holds. We numerically extract the long time slope
$\Gamma \Delta \mu$, and recover the weak power law dependence by
multiplying the correlation function by the inverse of the
exponential decay. The standard orthogonality effect is presented
in panels (a)-(b) for $\alpha=0.5$ and $\nu=0.3$, for which $\tilde \gamma=-0.038$.
When $\alpha=0$ and $\nu=0.8$ the anti-orthogonality effect clearly manifests itself
with $\tilde \gamma=0.12$  (c)-(d).
Interestingly, the correlation function at long times shows a
complicated behavior, more complex than that predicted in Eq.
(\ref{eq:Rfer}),  as evidenced by the mild deviations from strict
power law behavior displayed in Fig. \ref{Fig7}(d).

\begin{figure}
{\hbox{\epsfxsize=80mm \epsffile{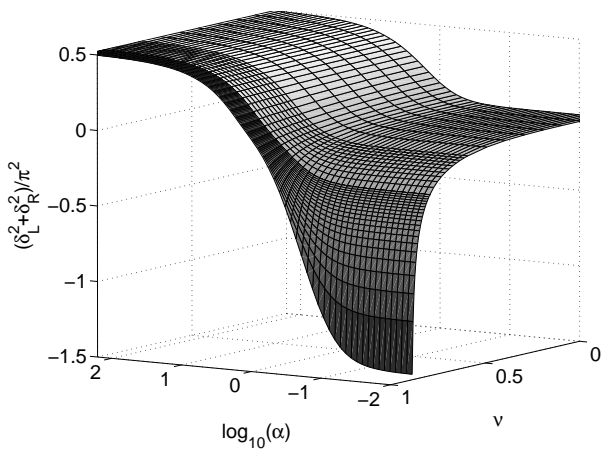}}} \caption{Map of
orthogonality [$(\delta_L^2+\delta_R^2)>0$] and anti-orthogonality
[$(\delta_L^2+\delta_R^2)<0$] behavior using expression
(\ref{eq:phaseC0}) with $\alpha=\alpha_1=\alpha_2$.}
 \label{Fig8}
\end{figure}

Fig. \ref{Fig8} presents an "orthogonality-anti-orthogonality" map as a
function of the diagonal ($\alpha$) and nondiagonal ($\nu$) couplings
using the general expressions of Eq. (\ref{eq:phaseC0})
with $\alpha=\alpha_1=\alpha_2$. We find that for large $\alpha$,
$-1/2<\tilde \gamma<0$, manifesting the standard orthogonality
effect. For large nondiagonal interactions typically
anti-orthogonality may be observed.



In addition to the long time exponential decay and anti-orthogonality
behavior, non-equilibrium dynamics may be reflected in the
appearance of complex power law exponents \cite{Ng}. When the
spin impurity is symmetrically coupled to the two leads
($\alpha_1=\alpha_2$), the phase shifts are complex conjugates,
see Eq. (\ref{eq:phaseC0}), and  $\tilde \gamma$ is always
real. This situation was discussed above. In contrast, {\it
asymmetric} systems may acquire a complex coefficient with $\tilde
\gamma= \tilde \gamma' +i\tilde \gamma''$, a direct outcome of a
non-equilibrium situation.

\begin{figure}
{\hbox{\epsfxsize=80mm \epsffile{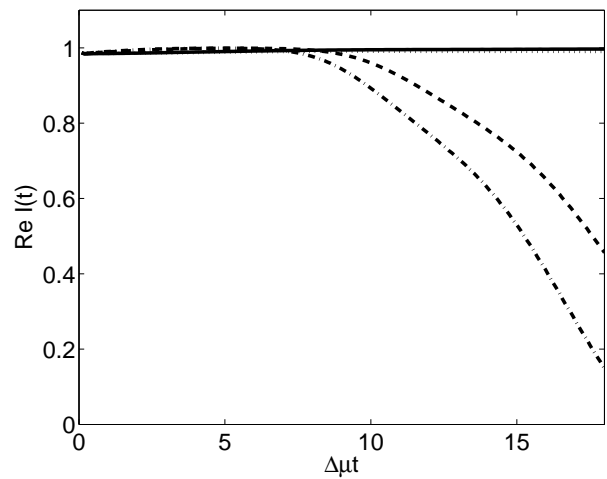}}}
\caption{Resolving the complex part of the power law exponent $\tilde \gamma$.
$\alpha_1=0$, $\alpha_2$=0 (full); $\alpha_1=0.1$, $\alpha_2$=0 (dashed);
$\alpha_1=0.2$, $\alpha_2$=0 (dashed-dotted);
$\alpha_1=0.2$, $\alpha_2$=0.2 (dotted);
$\Delta\mu=0.24$ and $\nu=0.95$ in all plots.
  }
 \label{FigC}
\end{figure}

The imaginary contribution to $\tilde \gamma$  is resolved in Fig.
\ref{FigC}.
Since at weak coupling the imaginary term $\tilde \gamma''$ is very small,
we investigate a strong coupling system with $\nu=0.95$.
Motivated by Eq. (\ref{eq:Rfer}), we
assume the generic form $C_f(t)=|C_f(t)|e^{i\epsilon t}
e^{i\tilde \gamma'' \ln t}$.
We numerically extract the
phase factor $\epsilon$, then plot the function
$I(t)=C_f(t)/|C_f(t)| e^{-i\epsilon t}$ for different diagonal
coupling strengths. As expected, in symmetric situations, $I(t)\sim 1$.
In contrast, asymmetric systems ($\alpha_1\neq \alpha_2$)
reveal an additional decaying contribution which is expected to
oscillate at longer times. We did not succeed in fitting
$I(t)$ to the stretched-oscillatory function $e^{i\tilde \gamma''
\ln t}$, indicating that at strong coupling the dynamics is
more involved. Finally, we note that, consistent with our
observations above, equilibrium effects dominate up to $\Delta\mu
t \sim 10 $. Only at longer times $I(t)$ begins to deviate
from unity due to the emerging influence of the imaginary term
$\tilde \gamma''$.

Though the imaginary term $\tilde \gamma''$ can strongly affect
the correlation function $C_f(t)$, (Fig. \ref{FigC}), its
practical contribution to the Golden Rule rate is small. We find
that for weak to intermediate coupling, $\tilde \gamma\ll
1$, leading to $I(t)\sim 1$. On the other hand, for strong
coupling, $\tilde \gamma''$ manifests itself only at long times
$\Delta\mu t >10$, when the correlation function
 has essentially decayed to zero.

\section{Relaxation}

\subsection{Qualitative discussion}

In this section we present calculations of the  nonadiabatic
relaxation rates $\Gamma_f^{\pm}(\omega)$ defined in  Eq. (\ref{eq:gaf}). %
The given physical model corresponds to
two states separated by an energy $\omega$ which depends on the
bare level splitting $B$ and on renormalizations arising from the
coupling to the leads. $\Gamma_f^+(\omega<0)$ corresponds to the
up-scattering rate describing transitions from the lower level to
the upper level, while $\Gamma_f^+(\omega>0)$ corresponds to
down-scattering. In equilibrium at $T=0$,
$\Gamma_f^+(\omega<0)$=0, i.e. there is no up-scattering. At
temperature $T>0$, the detailed balance relation of equilibrium
thermodynamics implies
$\Gamma_f^+(-\omega)/\Gamma_f^+(\omega)=e^{-\omega/T}$. In this
section we examine the rates in the non-equilibrium situation.
We show that the Gaussian form of the correlation
function $C_f(t)$  which occurs at strong coupling has important
consequences for the physics.

Before discussing our results in detail, we establish the relevant
energy scales. The general expression, Eq. (\ref{eq:gaf}), may be
written (neglecting overall factors) as
\begin{equation}
\Gamma_f^+(\omega)=\Re \int_0^\infty
\frac{dt}{\left(iDt\right)^{\beta_{eff}(t)}} e^{i\omega t-A(t)}.
\label{eq:Gamapprx}
\end{equation}
Here  $\beta_{eff}(t)$ is an effective exponent which changes from
the equilibrium power $\beta$, Eq. (\ref{eq:betadef}), to the non-equilibrium
value $\beta_{neq}$ [defined as $(-\gamma)$ in Eq. (\ref{eq:gamacof})],
as $\Delta \mu t$
changes from less than unity to much great then unity.
$\omega$ is the physical energy level difference, given by the sum of  $B$
[Eq. (\ref{eq:Hfermion})] and the level shift arising from the
system-bath coupling, and $D$ is an energy scale of the order of the Fermi sea bandwidth.

The naive assumption \cite{Aditisemi} is that the only important energy scale is
the relaxation rate given by the current flow across the system,
$\Gamma \Delta \mu$. In fact the numerical and analytical results presented in the previous
sections indicate that the situation is more subtle. At short times, $A(t) \approx \kappa (\Delta \mu t)^2$ whereas
at long times $A(t) \rightarrow \Gamma \Delta \mu t$.  The interplay of $\kappa$, which is proportional
to coupling strength $\nu^2$ at weak coupling but saturates at strong coupling [Eq. (\ref{eq:phineqkappa})],
and $\Gamma$ which is proportional
to $\nu^2$ at weak coupling but diverges at strong coupling [Eq. (\ref{eq:gamma})], gives a richer behavior.

Appendix E gives details of an asymptotic analysis of Eq. (\ref{eq:Gamapprx}). This analysis reveals that to discuss
the relaxation rate one should distinguish strong and weak coupling.
In the weak coupling limit, there are two relevant scales, $\Gamma \Delta \mu$
and $\Delta \mu$ (the latter multiplied by  various factors which are in practice fairly close to unity).
For $\omega>\Delta \mu$ we get the equilibrium down-scattering
rate; for $|\omega|<\Delta \mu$ we find a nontrivial approximately Lorentzian behavior, and for $\omega <-\Delta \mu$
we reproduce the $e^{-\frac{\omega}{\Delta\mu} \ln \frac{\omega}{\Delta\mu}}$
behavior found by Mitra et al. \cite{Aditisemi}.   Specifically,
\begin{eqnarray}
&=& {\tilde \Gamma}(1-\beta) \frac{\sin(\pi \beta)}{\omega} \left(\frac{\omega}{D}\right)^{\beta}; \hspace{0.1in} \omega> \Delta\mu
\nonumber\\
\Gamma_f^+(\omega)&=& \frac{\Gamma \Delta \mu}{\omega^2+\Gamma^2\Delta\mu^2} \hspace{0.1in}; -\frac{\Delta \mu}{\ln \Gamma}<\omega<\Delta \mu
\nonumber\\
&\sim& e^{-\frac{\omega}{\Delta\mu} \ln(\omega /\Delta \mu)}; \hspace{0.1in} \omega<-\frac{\Delta \mu}{\ln \Gamma}.
\label{eq:threeweak}
\end{eqnarray}
Here $\tilde \Gamma(x)$ is the complete Gamma function.
Note that the formulae match at $\omega \simeq \Delta\mu$ because in weak coupling
$\beta\simeq \Gamma\ll1$  leading to $\Gamma_f^+(\omega \sim \Delta\mu)\propto 1/\omega$.
%
In the strong coupling limit, two frequency scales turn out to be important:
$\sqrt{\kappa}\Delta \mu$ and $\Gamma \Delta \mu$.
We find
\begin{eqnarray}
&=&  {\tilde \Gamma}(1-\beta)  \frac{\sin(\pi \beta)}{\omega} \left(\frac{\omega}{D}\right)^{\beta}; \hspace{0.1in} \omega\gtrsim\sqrt{\kappa}\Delta \mu
\label{eq:Gamma-an1}\\
\Gamma_f^+(\omega)&=&
e^{-\frac{\omega^2}{4\kappa(\Delta \mu)^2}}
\frac{\tilde\Gamma[(1-\beta)/2]
\cos{\frac{\pi \beta}{2}} }{2\sqrt{\kappa} \Delta\mu}
\left(\frac{\sqrt{\kappa}\Delta \mu}{D}\right)^{\beta};
\nonumber\\
&&\hspace{1.1in}  |\omega|\lesssim\sqrt{\kappa}\Delta \mu
\label{eq:Gamma-an2}\\
&=& e^{-\frac{\omega^2}{4\kappa(\Delta \mu)^2}}
\frac{\cos(\pi\beta)}{2\sqrt{\kappa} \Delta \mu}\left(\frac{2\kappa \Delta \mu^2}{D\omega}\right)^\beta;
\nonumber \\
&&\hspace{1.0in} -\Gamma \Delta \mu\lesssim \omega\lesssim -\sqrt{\kappa}\Delta \mu
\label{eq:Gamma-an3}\\
&\sim& e^{-\frac{\omega}{\Delta\mu} \ln(\omega /\Delta \mu)}; \hspace{0.35in} \omega<-\Gamma \Delta \mu
\label{eq:Gamma-an4}
\end{eqnarray}
The Gaussian behavior found at intermediate frequency scales is a consequence of
the wide regime of $t^2$ behavior
found in the time evolution function, and may be roughly understood as the
Fourier transform of $e^{-\kappa (\Delta \mu t)^2}$
although as the results of Appendix E show, this argument must be treated with some care.

We call Eqs. (\ref{eq:Gamma-an2})-(\ref{eq:Gamma-an3})
 the "fermionic Marcus rate," the
analogue of the classical Marcus result for spin-boson systems
(\ref{eq:Gammab}), which holds in non-equilibrium situations at
strong coupling. This expression indicates that the voltage {\it
activates} the absorption rate, similarly to the role of
temperature in the bosonic case. The result differs from the
bosonic solution (Appendix B) in some important aspects: (i) In
the fermionic case the Gaussian decay is modified by a weak power
law term. (ii) For bosonic systems the activation factor depends
on the temperature as $\ln \Gamma_f \propto T^{-1}$,
while for fermionic systems we get a voltage squared activation, $\ln
\Gamma_f \propto \Delta\mu^{-2}$. Therefore, there is no
simple linear mapping between temperature and voltage drop in the
strong coupling regime. We note however that the classical Marcus
rate is applicable in the high temperature limit (see Appendix B),
while we typically assume here that $\Delta\mu\ll D$. Therefore,
in our system the energy window for reorganization processes is
the bias voltage, rather than the full bandwidth $D$. Thus, we may
interpret the $\kappa \Delta\mu^2$ factor in the denominator of
the Gaussian decay (\ref{eq:Gamma-an3}) as a reorganization energy
of the non-equilibrium fermionic system, $\lambda_f\sim\kappa
\Delta\mu$, multiplied by the driving force $F_f\equiv\Delta\mu$.
In contrast, in the equilibrium spin-boson model, reorganization
energies are of order of the cutoff frequency,
$\lambda_b\propto\omega_c$, and the driving force for absorption
processes is temperature $F_b\equiv T$. Qualitatively, both models
then recast to the familiar Marcus-like form, $\Gamma\propto
e^{-\omega^2/\lambda F}$ \cite{Marcus-comm}.
Further, both
the fermionic and the standard Marcus behaviors share similar
qualitative  features such as the existence of an inverted regime, as discussed in the next section.



\begin{figure}
{\hbox{\epsfxsize=90mm \epsffile{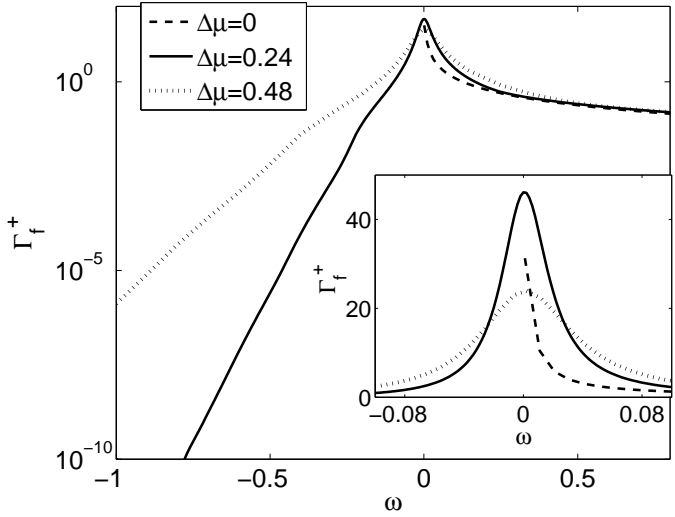}}} \caption{Golden Rule relaxation
rate calculated for the weak coupling limit $\nu=0.5$ ($\beta=0.043$, $\Gamma=0.08$),
$\Delta \mu=0$ (dashed line); $\Delta\mu=0.24$ (solid line);  $\Delta\mu=0.48$ (dotted line).
The inset presents an expanded view of low frequency regime.
}
\label{Relaxweakcoupling}
\end{figure}

\begin{figure}
{\hbox{\epsfxsize=90mm \epsffile{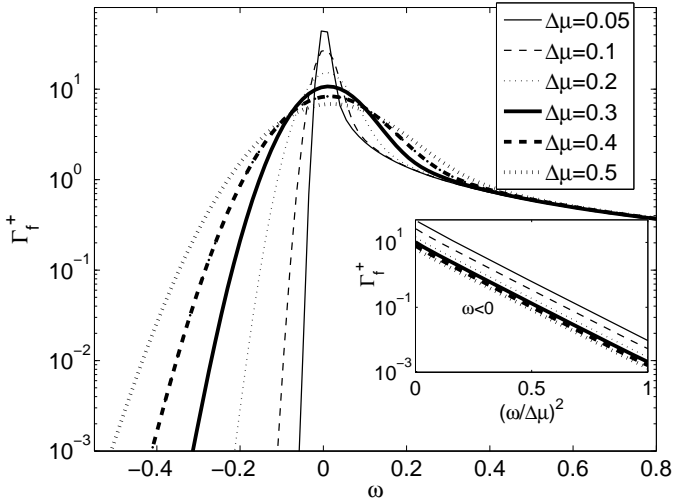}}} \caption{Fermi
Golden Rule rate, Eq. (\ref{eq:Gamapprx}), as a function of
frequency.
$\kappa=0.03$ and $\beta=0.13$ ($\nu=0.95$). Inset: Voltage activated excitation
rate
($\omega<0$).
 }
 \label{Fig9}
\end{figure}


\begin{figure}
\vspace*{3mm} {\hbox{\epsfxsize=80mm \epsffile{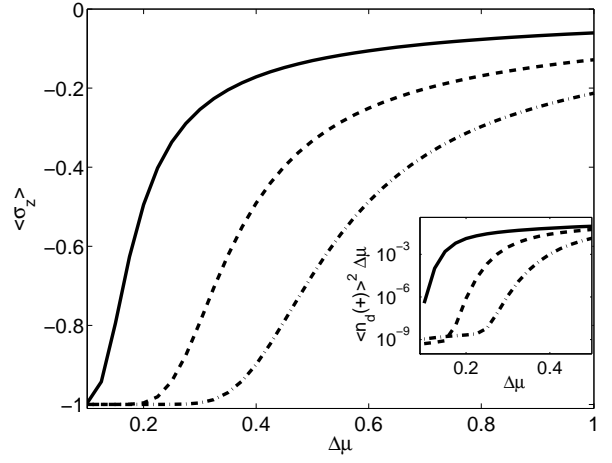}}} \caption{
Spin polarization as a function of potential bias $\Delta\mu$ for
different energy biases
 $\omega=0.1$ (full), $\omega=0.2$ (dashed), $\omega=0.3$ (dashed-dotted)
for $\kappa=0.03$, $\beta=0.13$ ($\nu=0.95$).
Inset: Mean field calculation of the current $I\propto \langle n_d(+)\rangle^2 \Delta\mu$ for the same frequencies as in the main plot.
}
 \label{Fig11}
\end{figure}


\subsection{Numerics: rates, population and current}

We numerically evaluate the integral (\ref{eq:Gamapprx}) using the
coefficients $\Gamma$, $\gamma$, $\beta$ and $\kappa$ as determined by the coupling
strengths, Eqs. (\ref{eq:gamma}), (\ref{eq:gamacof})
(\ref{eq:phieq}) and (\ref{eq:phineqkappa})
respectively. For convenience, we disregard the multiplicative
factor $\Delta^2/2$.

The main panel of Fig. \ref{Relaxweakcoupling} shows on a
semi-logarithmic scale the relaxation rate computed numerically
for the relatively weak coupling $\nu=0.5$ ($\beta=0.043$,
$\Gamma=0.08$, non-equilibrium exponent
$\beta_{neq}=-\gamma=0.013$) and two choices of chemical potential,
$\Delta\mu=0.24$ and $\Delta\mu=0.48$.
Also shown as the dashed line is the $T=0$  equilibrium result.
The inset shows an expanded view of the
small frequency regime, demonstrating the Lorentzian behavior.
We clearly observe the three regimes as discussed in Eq. (\ref{eq:threeweak}):
For small frequencies (large bias voltage)
the spin levels are approximately degenerate,
and the rates are symmetric around $\omega=0$ (inset).
In the opposite  $|\omega|>\Delta\mu$ limit, the absorption rate
is practically  zero, while the emission rate
approaches the equilibrium limit.
In between, a voltage activated excitation behavior is revealed.

We analyze next the strong coupling limit. Fig. \ref{Fig9} shows
that the excitation process is activated by a finite potential
difference as prescribed by Eq. (\ref{eq:Gamma-an3}). More
quantitatively, the inset verifies that the relationship  $\ln
\Gamma_f^+(\omega<0)\propto -(\omega^2/\Delta \mu^2)$ holds.
Similar to classical bosonic Marcus rate \cite{Marcus}, an
inverted regime appears for the fermionic system. However, in the
present case the rate in the inverted regime decays {\it weakly}
as a power law rather than as a Gaussian. At large frequencies,
$\omega\gg \Delta\mu$, equilibrium behavior is observed where
$\Gamma_f^+(\omega<0)$ approached zero, and $\Gamma_f^+(\omega>0)$
becomes insensitive to voltage.
We have also calculated the Golden Rule rate using
the {\it numerical} correlation function
(depicted e.g. in Figs. \ref{Fig1} and \ref{Fig4}),
instead of the approximate analytical function in (\ref{eq:Gamapprx}),
and find that the results agree perfectly.

We now turn to a study of the spin polarization. In the incoherent tunneling
regime, for small tunneling parameter $\Delta$, the populations of
the two levels obey a Markovian balance equation
\beq \dot P_{+}= \Gamma_f^-P_-- \Gamma_f^+P_+;
\,\,\,\,\,  P_-+P_+=1, \eeq
with the absorption and emission rates given by Eq.
(\ref{eq:Gamapprx}). 
The polarization $\langle \sigma_z \rangle \equiv P_+-P_-=
\frac{\Gamma_f^--\Gamma_f^+}{\Gamma_f^{+}+\Gamma_f^{-}}$, shown in
Fig. \ref{Fig11}, manifests a transition from a fully polarized
system $\langle \sigma_z \rangle \sim -1$ to an unpolarized system
$\langle \sigma_z \rangle \sim 0$ as $\Delta \mu$ is increased.
Typically, we find that the crossover takes place when the energy
bias $\omega$ becomes comparable to the bias voltage. While at
high frequencies, $|\omega| \gg \Delta\mu$,
$\Gamma_f^+(\omega<0)\sim0$, leading to full polarization, at very
large bias the emission and absorption rates are comparable,
resulting in equal population of the two levels and zero
polarization. The  Gaussian activation term in Eq.
(\ref{eq:Gamma-an3}) is therefore reflected in the enhancement of
polarization with bias voltage.

It is also interesting to note that the electron
current through the system, calculated at the level of mean field theory,
$I\propto \Delta\mu \langle n_d(+) \rangle^2$, is strongly suppressed
for weak bias, $\Delta\mu<\omega$, see inset of Fig. \ref{Fig11}.
In contrast, for very large bias, $\langle n_d(+) \rangle=1/2$, and the current
increases linearly with $\Delta\mu$.
Therefore, it is the intermediate regime of $\omega\sim\Delta\mu$
that manifests prominent nonlinear current-voltage characteristics,
emerging due to the interplay between the Gaussian relaxation and the power-law dynamics.
We can compare our results to the weak coupling Bloch-type rate equations of
Gurvitz {\it et al.} \cite{Gurvitz} which yield
$\langle \sigma_z\rangle =0$ at long enough times,
independent of voltage drop and energy bias.
In contrast, Fig. \ref{Fig11}
reveals a  rich dynamics in the strong coupling regime with a prominent dependence
on system energetics and the non-equilibrium conditions.


\section{Beyond ${\cal O}(\Delta^2)$: Coulomb Gas behavior}

\begin{figure}
{\hbox{\epsfxsize=70mm \epsffile{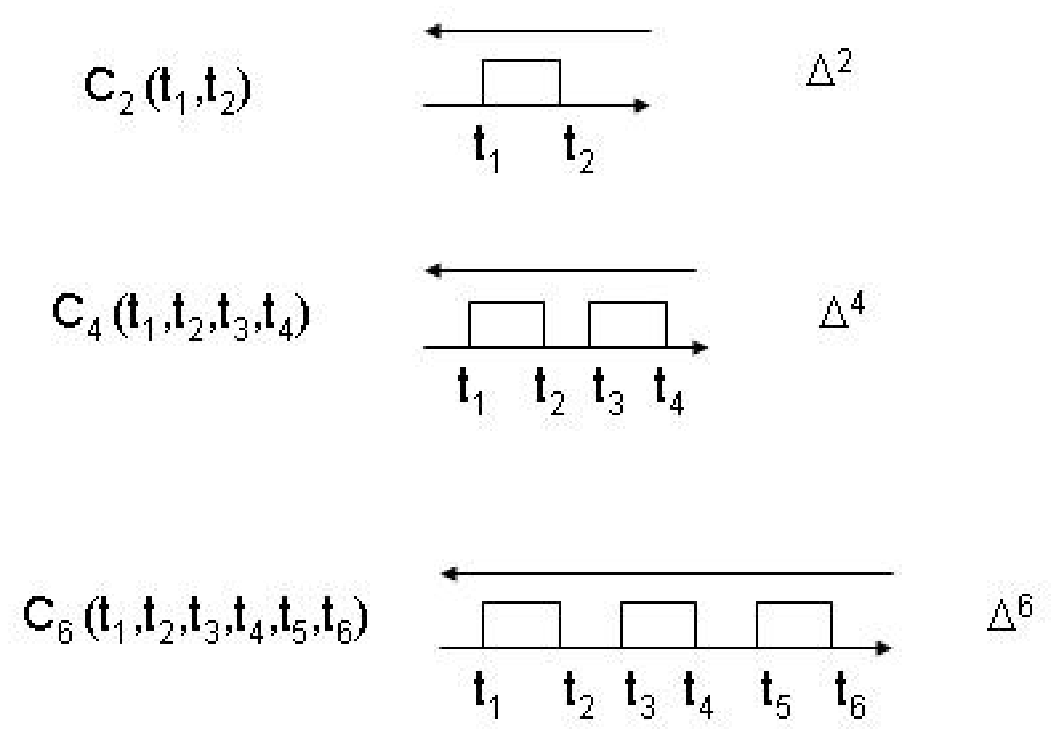}}}
\vspace*{-6mm}
\caption{Schematic representation of spin-flip evens on the Keldysh contour.
Plotted are examples for particular two, four and six spin-flip processes,
respectively.}
 \label{scheme}
\end{figure}

In this section we discuss a crucial ingredient of the physics of our
model that allows for a description beyond the Golden Rule [${\cal O}(\Delta^2)$]
level.
A formally exact solution for the impurity spin problem
(\ref{eq:Hfermion}) can be written by a power series in the tunneling
matrix element $\Delta$ \cite{Chang}.
Here, we restrict ourselves to an exact numerical investigation
of the electronic correlation functions that appear in this power series
to see if the usual "Coulomb gas" behavior is observed even when
the system is out-of-equilibrium.
In particular, the reduced density matrix of the spin impurity $\rho_d(t)$
is given by
\beq \rho_d(t)={\rm Tr} \left[ e^{-iHt}\rho(0)e^{iHt}\right], \eeq
with forward and backward time evolution branches.
Here $\rho$ is the total density matrix,
and the trace is performed over the reservoir electronic states.
We decompose the propagators, including all spin-flip events along the
time ordered contour, and obtain,
e.g. for the spin up population \cite{Leggett,Chang},
 \bea &&\langle +|\rho_d(t) |
+\rangle = \sum_{k=0}^{\infty}\sum_{j=0}^{\infty} (-1)^{(k+j)}
\left(\frac{\Delta}{2}\right)^{2k+2j}
\nonumber\\
&&\times
 \int_0^{t}dt_1 \ldots \int_{0}^{t_{2k-1}}dt_{2k}\int_{0}^{t}ds_1 \ldots
\int_{0}^{s_{2j-1}}ds_{2j}
\nonumber\\
&&\times {\rm Tr}\bigg[ \bigg(e^{iH_+s_{2j}}   e^{-iH_-s_{2j}}\bigg) \ldots
\bigg(e^{iH_-s_{1}}   e^{-iH_+s_{1}}\bigg)
\nonumber\\
&&\times
\bigg(e^{iH+t_{1}}   e^{-iH_-t_{1}}\bigg) \ldots
\bigg(e^{iH_-t_{2k}}   e^{-iH_+t_{2k}}\bigg)
 \bigg].
\nonumber\\
\label{eq:exact}
\eea
Here $|\pm \rangle$ are the up and down spin states,
and $H_{\pm}$ is defined in Eq. (\ref{eq:Hpm}).
This expression was derived assuming that at
$t=0$ the spin is in the pure
state $|+\rangle$, and the (isolated) reservoirs are in their
respective ground states ($T=0$).
It can be easily generalized to describe other initial conditions.
Each time variable in Eq. (\ref{eq:exact}) marks a particular
spin-flip event. While the second order correlation
function couples nearest neighbor events only, higher order
correlations couple distant spin-flips, yielding a multiparticle
interaction term.
In equilibrium, this interaction can be exactly written
in terms of pair-wise contributions,
${\rm Tr} [...] \propto \exp\left[ \sum_{k<j}(-1)^{k+j}\Phi_f(t_k-t_j) \right]$,
significantly simplifying the computational problem.
This is the celebrated Anderson-Yuval-Hamann (AYH) result,
which leads to the interpretation of the Kondo problem
as a one-dimensional Coulomb gas system \cite{AYH}.
In contrast, in the general non-equilibrium case,
the exact structure of the interaction
is not known for all times,
and it is not clear whether higher order correlations can be exactly
decomposed into pair-wise contributions \cite{Aditicoul}.

Using the numerical technique discusses in section IV, we can exactly
calculate, term by term, the correlation functions in Eq. (\ref{eq:exact}).
Specifically, we study three examples of the processes of the order of $\Delta^2$, $\Delta^4$ and
$\Delta^6$, depicted schematically in Fig. \ref{scheme},
\bea &&C_{2}(t_1,t_2)=\langle e^{iH_-t_1} e^{iH_+(t_2-t_1)}
 e^{iH_-(t-t_2)} e^{-iH_-t} \rangle
\nonumber\\
&&C_4(t_1,t_2,t_3,t_4)=
\nonumber\\
&&\langle e^{iH_-t_1} e^{iH_+(t_2-t_1)}
e^{iH_-(t_3-t_2)}
 e^{iH_+(t_4-t_3)}   e^{iH_-(t-t_4)}  e^{-iH_-t} \rangle
\nonumber\\
&&C_6(t_1,t_2,t_3,t_4,t_5,t_6)=
\nonumber\\
&&\langle e^{iH_-t_1}
e^{iH_+(t_2-t_1)}  e^{iH_-(t_3-t_2)}
 e^{iH_+(t_4-t_3)}   e^{iH_-(t_5-t_4)}
\nonumber\\
&&\times
 e^{iH_+(t_6-t_5)}  e^{iH_-(t-t_6)}   e^{-iH_-t} \rangle,
\label{eq:correx}
\eea
and compare the results to the Coulomb gas expressions
\bea
&&\tilde C_2(t_1,t_2)\equiv C_2(t_1,t_2)=C_2(t_2-t_1),
\nonumber\\
&&\tilde C_{4}(t_1...t_4)=
\nonumber\\
&&\frac{C_2(t_2-t_1)C_2(t_4-t_1)C_2(t_3-t_2)C_2(t_4-t_3)}
{C_2(t_3-t_1)C_2(t_4-t_2)},
\nonumber\\
&&\tilde C_6(t_1...t_6)= \tilde C_4(t_1..t_4) \times
\nonumber\\
&&
\frac{C_2(t_6-t_1)C_2(t_5-t_2) C_2(t_6-t_3) C_2(t_5-t_4)C_2(t_6-t_5)}
{C_2(t_5-t_1)C_2(t_6-t_2)C_2(t_5-t_3)C_2(t_6-t_4)}.
\nonumber\\
\label{eq:corrapp} \eea
In particular, our calculations were performed  assuming a regular
interval $\tau$ between spin flips.

\begin{figure}
{\hbox{\epsfxsize=80mm \epsffile{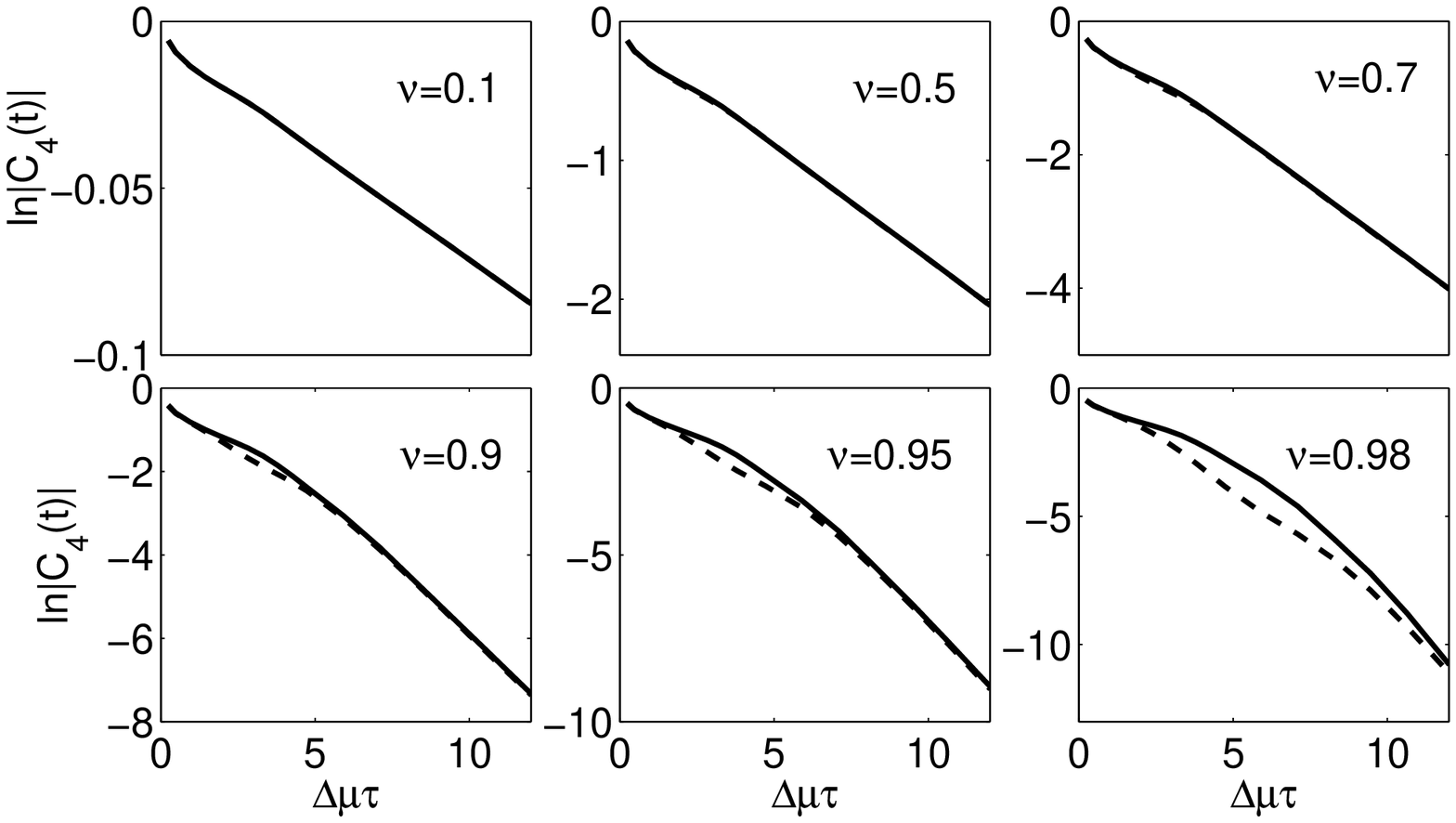}}} \caption{
Testing the Coulomb gas picture for out-of-equilibrium situations.
Comparison between the exact fourth order correlation function
$C_4(t_1,t_2,t_3,t_4)$, Eq. (\ref{eq:correx}) (full),
and the pair-wise approximation $\tilde C_4(t_1,t_2,t_3,t_4)$, Eq.
(\ref{eq:corrapp}) (dashed).
$\tau=t_{i+1}-t_{i}$ is the distance
between spin-flip events. All other parameters are the same as in
Fig. \ref{Fig2}.}
 \label{Fig12}
\end{figure}

\begin{figure}
{\hbox{\epsfxsize=80mm \epsffile{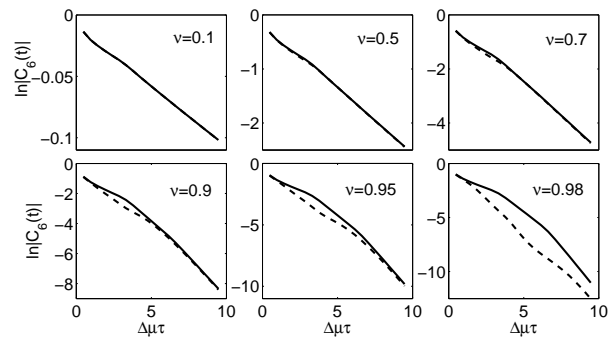}}} \caption{
Testing the Coulomb gas picture for out-of-equilibrium situations.
Comparison between the exact expression sixth order correlation function
$C_6(t_1,t_2,t_3,t_4,t_5,t_6)$, Eq. (\ref{eq:correx}) (full), and
the pair-wise approximation $\tilde C_6(t_1,t_2,t_3,t_4,t_5,t_6)$,
Eq. (\ref{eq:corrapp}) (dashed).
$\tau=t_{i+1}-t_{i}$ is the
distance between spin-flip events. All other parameters are the
same as in Fig. \ref{Fig2}.}
 \label{Fig13}
\end{figure}

We have robustly checked that the AYH decomposition \cite{AYH}
holds precisely at all times (greater than $D\tau>1$) and coupling strengths in
equilibrium, as well as for all times out-of-equilibrium, for weak to
intermediate coupling strengths, see Figs. \ref{Fig12}-\ref{Fig13}.
Interestingly, the AYH decomposition breaks down for the strong coupling out-of-equilibrium situation, precisely
in the time window where $C_f(t)$ shows a broad Gaussian decay with time, as depicted
in Fig. \ref{Fig2}.
Even in this regime, the pair-wise AYH decomposition
holds asymptotically for long and short times.

An important outcome of this observation is that the AYH Coulomb gas expression \cite{AYH},
which is exact in equilibrium, cannot be justified for intermediate times 
$\Delta\mu\sim1-10$ for strong coupling to the leads in the out-of-equilibrium situation.
This is because at strong coupling
the effective short time behavior (which cannot be described
by the Coulomb gas picture) practically extends to longer times of
order $\Delta\mu t \sim 1-10$.
The Coulomb gas expression still holds for weak to intermediate coupling strength and at long times.
This investigation lays the groundwork for an exact evaluation of the spin dynamics 
via path integral techniques valid even when the Coulomb gas decomposition 
does not hold.  
This work will be reported in a future publication \cite{Next}.


\section{Summary}

In this paper we have undertaken a detailed study of the non-equilibrium
dynamics of a small quantum system coupled to two electronic leads.
This problem is of great interest for understanding dissipative effects
in prospective single molecules devices. The model studied here is generic
enough to capture range of  relevant relaxation motifs,
while being simple enough for detailed investigation.
Our analysis combines analytical results with exact numerics,
rendering a detailed and  clear picture of the dissipative behavior on
{\it all time scales} for {\it arbitrary strong coupling to the leads}.

While previous works have studied the non-equilibrium dynamics in the
long time limit \cite{Ng,Braunecker1,Aditispin}, we have provided new
information in the intermediate time domain, where exact analytical results
are not available.
In the nonadiabatic limit for strong system-lead coupling
we have discovered a new non-equilibrium regime with a Marcus-like
spin relaxation rate.
Here, while the non-equilibrium dynamics is qualitatively similar to the
equilibrium dynamics at a finite temperature, the analogy is not complete.
In particular, a simple linear mapping between temperature and bias voltage
does not exist,
in contrast to the electrically damped harmonic oscillator
model \cite{Martin,Martin03}.
The Marcus-like relaxation rate  exhibits highly
 nonlinear current-voltage (I-V) characteristics:
The current is practically suppressed at small
bias voltage,  is strongly enhanced at intermediate bias (of
the order of the energy difference between spin levels $B$),
while for large bias linear I-V behavior emerges.

In the long time limit a non-equilibrium situation generates complex scattering
phase shifts which are
reflected in the dynamics through different effects:
(i) onset of an exponential decay for the spin polarization,
(ii) appearance of a power law term in the relaxation dynamics,
with a {\it complex} exponent, and
(iii) the possible existence of an anti-orthogonality-regime.
The effects presented in this paper are not limited to the
specific model utilized here, but can be rederived for other
systems, e.g. the resonant level model of Appendix A,
where the polarization of a spin impurity couples to the resonant
level occupancy \cite{Aditispin}.

Going beyond the nonadiabatic limit, we have studied multiple
spin-flip events with the aid of exact numerical calculations.
Interestingly, we have found that the Anderson-Yuval-Hamann
treatment of the equilibrium Kondo effect
\cite{AYH} can be extended to the out-of-equilibrium regime, but
only for weak to intermediate system-bath couplings does the standard
pair-wise Coulomb gas behavior hold qualitatively for all timescales \cite{Aditicoul}.
Deviations occur precisely in time intervals where the Gaussian
decay of the correlation function $C_f(t)$ is prominent.

Several future directions are worthy of investigation. First, we have
restricted ourselves in this work to zero temperature.
Including the effect of finite temperature is straightforward both analytically
and numerically \cite{Temp}. In particular, the mapping
$t \rightarrow {\rm tanh}(\pi k_bT t)$ \cite{Braunecker2} transforms
all analytical expressions to those valid at finite (but low $k_BT<\Delta\mu$)
temperatures. Similarly, the numerical approach of section IV.A
may be generalized to arbitrary temperatures.
The simple model studied here can be extended in several important ways, including
coupling of the quantum subsystem to vibrational degrees of freedom.

Lastly, we have delineated the precise set of regimes where the
standard Anderson-Yuval-Hamann Coulomb gas behavior is quantitatively accurate.
This lays the groundwork for future exact numerical studies of the spin dynamics
for the models discussed here. In particular, standard influence functional
methodology may be directly applied in regimes where pair-wise Coulomb gas
behavior is exhibited \cite{Aditicoul}.
In regimes where deviations exist, numerically exact Monte Carlo without the pair-wise 
assumption may be performed.
Both of these approaches are currently being pursued.

\acknowledgments This work was supported by NSF (NIRT)-0210426 (DS and DHR) and DMR-0705847 (AJM).
The authors acknowledge M. S. Hybertsen and A. Mitra for fruitful discussions.

\renewcommand{\theequation}{A\arabic{equation}}
\setcounter{equation}{0}  
\section*{\label{Resonant}APPENDIX A: The Spin-resonant-level model}

We present here a variant of the model system (\ref{eq:Hfermion})
leading to dynamics analogous to Eqs. (\ref{eq:threeweak})-(\ref{eq:Gamma-an4}).
The model was  presented in Ref. \cite{Aditispin}
for the analysis of the generalized fluctuation-dissipation
relation for out-of-equilibrium systems. It describes a spin
system coupled to a spinless resonant level (creation operator
$d^{\dagger}$), which is itself coupled to two electron baths
$n=L,R$,
\bea H&=&H_S+H_{B}^{(f)}+H_{SB}^{(f)};
\nonumber\\
H_S&=&\frac{B}{2}\sigma_z+\frac{\Delta}{2}\sigma_x,
\nonumber\\
H_B^{(f)}&=&\sum_{k,n}\epsilon_ka_{k,n}^{\dagger}a_{k,n}
+\sum_{k,n}V_{k,n}\left( a_{k,n}^{\dagger}d+d^{\dagger} a_{k,n}
\right),
\nonumber\\
H_{SB}^{(f)}&=&J_zd^{\dagger} d \frac{(I+\sigma_z)}{2}. \label{eq:Hreson} \eea

Here $B$ and $\Delta$ are the spin parameters, describing the
energy gap and the tunneling splitting respectively. $J_z$
reflects the strength of system-bath interaction, and $V_{k,n}$
is the coupling element of the resonant level to the $n$-th electronic
reservoir.
$I$ is the identity operator.
The relation of this model to the generic
Hamiltonian (\ref{eq:Hfermion}) is revealed by diagonalizing
$H_B^{(f)}$, and rewriting Eq. (\ref{eq:Hreson}) in terms of the
new operators as follows,
\bea H_B^{(f)}&=&\sum_{k,n}\epsilon_{k}c_{k,n}^{\dagger}c_{k,n},
\nonumber\\
H_{SB}^{(f)}&=&
\frac{J_z}{2}(I+\sigma_z)\sum_{k,n,k',n'}\nu_{k,n}^{*}\nu_{k',n'}c_{k,n}^{\dagger}c_{k',n'},
\nonumber\\
\nonumber\\
a_{k,n}&=&\sum_{k',n'} \eta_{k,n;k',n'}c_{k',n'} \,\,\,\, ; \,\,\,\
d=\sum_{k,n}\nu_{k,n}c_{k,n}. \label{eq:Hd} \eea
with $n,n'=L,R$. The coefficients $\nu_{k,n}$ and $\eta_{k,n,k',m}$
are given by \cite{Aditispin}
\bea
\nu_{k,n}&=&\frac{V_{k,n}}{\epsilon_k-\sum_{k',m}\frac{V^2_{k',m}}{\epsilon_k-\epsilon_{k'}-i\delta}},
\nonumber\\
\eta_{k,n;k',n'}&=&\delta_{k,k'}\delta_{n,n'}-\frac{V_{k,n}\nu_{k',n'}}{\epsilon_k-\epsilon_k'+i\delta},
\eea
where $\delta$ goes asymptotically to zero. Next we assume that
the resonant level-lead coupling is a constant, independent of
momentum. The phase shifts, complex numbers in
non-equilibrium situations, then become
\bea \delta_L&=&{\rm atan}
\frac{\lambda\sin^2(\theta)}{1-i\lambda\cos^2(\theta)},
\nonumber\\
\delta_R&=&{\rm atan}
\frac{\lambda\cos^2(\theta)}{1+i\lambda\sin^2(\theta)}.
\label{eq:phaseCres} \eea
Here $\lambda$ is a dimensionless coupling strength and
$\tan(\theta)$ determines the asymmetry with respect to coupling to the $L$ and $R$ sides
\bea \lambda&=&\frac{2J_z}{\Gamma_L+\Gamma_R}; \,\,\,\,
\theta=\rm{atan}\left(\sqrt{{\Gamma_L}/{\Gamma_R}}\right),
\eea
$\Gamma_n=2\pi V_n^2\rho$ is the hybridization of the resonant
level with the $n$-th reservoir and $\rho$ is the reservoirs [Eq.
(\ref{eq:Hreson})] density of states. When the system is
symmetric, $\Gamma_L=\Gamma_R$, $\theta=\pi/4$, and we obtain the
following relations
\bea
\delta_{eq}&=&\delta_L+\delta_R={\rm atan }(\lambda),
\nonumber\\
|\delta_L''-\delta_R''|&=&\frac{1}{2}\ln(1+\lambda^2),
\nonumber\\
 \delta_L^2+\delta_R^2&=& \frac{1}{2}\left[ {\rm atan}^2 \lambda -\frac{1}{4}\ln^2(1+\lambda^2)  \right],
\nonumber\\
2\delta_L\delta_R&=&
\frac{1}{2}\left[{\rm atan}^2 \lambda  +\frac{1}{4}\ln^2 (1+\lambda^2) \right].
\eea
At weak coupling the correlation function can be
derived using the cumulant expansion, as done through Eqs.
(\ref{eq:expansion})-(\ref{eq:cases}),
\bea
\Phi_f(t)&=& \frac{\lambda^2}{\pi^2} \ln(1+iDt) +iE_s t
\nonumber\\
&+& \frac{\lambda^2}{2\pi^2}\Delta\mu t  \bigg [{\rm Si}(\Delta\mu t) -\frac{1-\cos(\Delta\mu t)}{\Delta \mu t}\bigg ]
\nonumber\\
&-& \frac{\lambda^2}{2\pi^2} [\gamma_e+\ln(\Delta\mu t)- {\rm Ci}(\Delta\mu t)].
\eea
with $E_s=D\lambda/\pi$. By following the derivation which leads to
Eqs. (\ref{eq:gamma}), (\ref{eq:gamacof}), (\ref{eq:phieq}) and
(\ref{eq:phineqkappa}) for the present case, we obtain the spin-resonant level
correlation function at strong coupling,
\bea
C_{f}(t) \sim
\begin{cases}
(Dt)^{-\beta} & \Delta\mu t \ll 1  \\
(Dt)^{-\beta}e^{-\kappa \Delta\mu^2 t^2} & \Delta\mu t \sim 1-10  \\
(Dt)^{-\beta}(\Delta\mu t)^{\gamma}e^{-\Gamma\Delta\mu t}  & \Delta\mu t \gg 1, \\
\end{cases}
\eea
with the coefficients
\bea \beta&=& {\rm atan}^2(\lambda)/\pi^2,
\nonumber\\
\Gamma&=& \frac{1}{4 \pi} \ln(1+\lambda^2),
\\
\kappa&=&\frac{ {\rm atan}^2( \lambda) }{8\pi^2},
\nonumber\\
\gamma&=& \frac{1}{2\pi^2}\left[{\rm atan}^2 \lambda  +\frac{1}{4}\ln^2 (1+\lambda^2) \right].
\nonumber
\eea
We note that the orthogonality-anti-orthogonality transition takes place
when the exponential
$(\delta_L^2+\delta_R^2)$ changes sign, at $\frac{1}{2}\ln(1+\lambda^2)={\rm atan} (\lambda)$.

\renewcommand{\theequation}{B\arabic{equation}}
\setcounter{equation}{0}  
\section*{\label{Marcus} APPENDIX B:  Derivation of the classical Marcus rate in the spin-boson model}
In this Appendix we derive the classical Marcus behavior of a two-level system
coupled to an oscillator bath, and compare the result to the non-equilibrium
Marcus-like behavior found in the main text.
The classical Marcus result \cite{Marcus} emerges in the high temperature limit of the
asymmetric spin-boson model in the nonadiabatic regime \cite{WeissBook}.
The Hamiltonian is given by
\beq
H=H_S+H_B^{(b)}+H_{SB}^{(b)},
\label{eq:Htotal}
\eeq
where the spin system $H_S$ includes a two-level system (TLS) with
a bare tunneling amplitude $\Delta$ and a level splitting $B$.
The reservoir $H_B^{(b)}$ includes a set of independent harmonic oscillators, and the system-bath interaction $H_{SB}^{(b)}$ is bilinear in the reservoir coordinates and the
spin polarization
\bea
H_S&=&\frac{B}{2}\sigma_z+\frac{\Delta}{2} \sigma_x,
\nonumber\\
H_B^{(b)}&=&\sum_{j}\omega_j b_j^{\dagger}b_j,
\nonumber\\
H_{SB}^{(b)}&=&\sum_{j}\frac{\lambda_j}{2}(b_j^{\dagger}+b_j)\sigma_z.
\label{eq:Hboson}
\eea
Here $b_j^{\dagger}, b_j$ are bosonic creation and annihilation operators,
respectively. In the nonadiabatic regime the excitation rate can be calculated within Fermi's Golden Rule as \cite{Mahan}
\bea
\Gamma_b^{\pm}&=&{\left(\frac{\Delta}{2}\right)^2}\int_{-\infty}^{\infty}
dt e^{\pm i B t} C_b(t) ;\,\,\, \,\,\,\, C_b(t)=e^{-\Phi_b(t)};
\nonumber\\ \Phi_b(t)&=&\sum_j\frac{\lambda_j^2}{\omega_j^2}\left[(1+2n_j)-(1+n_j)e^{-i\omega_jt}-n_je^{i\omega_jt}\right]
\nonumber\\
&=&\int_0^{\infty}\frac{d\omega}{\pi} \frac{J(\omega)}{\omega^2}
\left[
1-e^{-i\omega t} +4n(\omega)\sin^2\frac{\omega t}{2}
\right],
\nonumber\\
\label{eq:Rbos}
\eea
with spectral function $J(\omega)= \pi \sum_{j} \lambda_j^2\delta(\omega-\omega_j)$.
Here $n(\omega)=[e^{\omega/k_BT}-1]^{-1}$ is the Bose-Einstein
distribution function with $T$ as the temperature of the bosonic
reservoir, $k_B$ is the Boltzmann constant.
The case relevant to the present paper is the Ohmic spectral density: a continuum of bosons
with $J(\omega)=2\pi \alpha \omega$ at low frequencies and $J(\omega) \rightarrow 0$
for $\omega$ greater than a cutoff scale $\omega_c$. A conventional choice  is
\bea
J(\omega)=2\pi \alpha \omega e^{-\omega/\omega_c},
\eea
but most of the results do not depend on this choice.
It is useful to decompose  $\Phi_b$ into two contributions
\bea
\Phi_b(t)= \Phi_1(\omega_ct) + \Phi_2(\omega_ct; Tt).
\eea
Here
\bea
\Phi_1=\int_0^{\infty} \frac{d\omega}{\pi} \frac{J(\omega)}{\omega^2}
\left(1-e^{-i\omega t} \right),
\nonumber\\
\Phi_2=4 \int_0^{\infty} \frac{d \omega}{\pi} \frac{J(\omega)}{\omega^2}
\frac{\sin^2 \frac{\omega t}{2} }{ e^{\omega/k_BT}-1}.
\label{eq:phi1phi2}
\eea
The first term gives the bosonic analogue of the zero temperature power
law dependence $\ln(1+iDt)$  found in the main text,
\bea
\Phi_1(\omega_ct \ll 1)&\sim& i\lambda t; \,\,\,  \lambda\equiv \sum_{j}\lambda_j^2 /\omega_j
=2 \alpha \omega_c,
\nonumber\\
\Phi_1(\omega_ct\gg1)&\sim&  2 \alpha \ln(i \omega_c t),
\label{eq:phi1Marcus}
\eea
where $\lambda$ is the solvent reorganization energy.
The second term gives the analogue of the exponential/Gaussian behavior.
For $k_BT>\omega_c$, the canonical  Marcus result is obtained. In this limit one
approximates $e^{\omega/k_BT}-1 \rightarrow \frac{\omega}{k_BT}$ leading to
\bea
\Phi_2^{Marcus}=4k_BT \int_0^{\infty} \frac{d\omega}{\pi} \frac{J(\omega)}{\omega}
\frac{\sin^2 \frac{\omega t}{2}}{\omega^2}.
\eea
At $\omega_ct\ll 1$ we approximate $\sin^2 \frac{\omega t}{2} \sim \omega^2 t^2 /4$
and obtain
\beq
\Phi_2^{Marcus}(\omega_ct\ll 1)= k_BT\lambda t^2.
\label{eq:phic}
\eeq
For $\omega_c t \gg 1$ we may set $J(\omega)=2 \pi \alpha \omega$
and get a linear behavior,
\bea
\Phi_2^{Marcus}(\omega_ct\gg 1)= 2 \pi \alpha k_BT t.
\label{eq:phi2Marcus}
\eea
The crossover scale is $t^*=1/\omega_c$, with the value
$\Phi_2^{Marcus}(t^*)\approx 2 \alpha k_BT/\omega_c$.
Thus, for any $\alpha$, at large enough $k_BT/\omega_c$, $\Phi_2$ becomes large enough that a
Gaussian relaxation results.

The analogy between the high temperature Marcus behavior
and the results we have found in  the non-equilibrium fermionic model is not complete, since in the
latter case we typically assume that the electron bands are wide
relative to the potential bias. We therefore consider next
the analogous equilibrium  limit of $k_BT< \omega_c$.
In this case the short time limit ($\sin(\omega t/2) \rightarrow \omega t/2$ )
of Eq. (\ref{eq:phi1phi2}) obeys
\bea
\Phi_2^T(\omega_ct\ll1) &=& 2 \alpha t^2
\int_0^{\infty} d\omega \frac{\omega e^{-\omega/\omega_c}}{e^{\omega/k_BT}-1}
\nonumber\\
&\sim& \frac{\alpha t^2k_B^2 T^2 \pi^2}{3},
\eea
while the long time limit reduces to (\ref{eq:phi2Marcus}).
We thus obtain a short time $t^2$ and a long time $t$-linear behavior. The crossover occurs at
$t^*\sim 1/k_BT$ and $\Phi_2^T(t^*)=\pi^2 \alpha/3 \sim 3 \alpha$.
Thus, for $\alpha$ much smaller than 1, the relaxation integrals are dominated
by the long time region where $\Phi \sim t$, leading to an exponential relaxation.
For $\alpha >0.5$ however, the power law prefactor  ensures that the integral
is dominated by short times, of order $\omega_c^{-1}$, so that the frequency dependence
is significant only on the scale of the cutoff scale $\omega_c$.

We compare next this behavior to the non-equilibrium Marcus-like behavior
found in the main text.
The mathematical essence of the non-equilibrium result is that the time decay function
behaves as $\Phi_f\sim \kappa (\Delta\mu t)^2$ at short times, and as
$\Phi_f\sim\Gamma \Delta\mu t$
at long times; the crossover between these two regimes occurs at
$t^* \sim \Gamma/\kappa \Delta\mu$ and the value at $t^*$ is
$\Phi_f(t^*)\sim \Gamma^2/\kappa$. In the weak coupling limit $\Gamma \sim \kappa$,
$t^*\sim 1/\Delta\mu$ and $\Phi_f(t^*)\sim \Gamma^2/\kappa \sim \nu^2 \ll 1$, so
the $t^2$  behavior is not important for the relaxation rates. However, in the strong
coupling limit, $\Gamma/\kappa \gg 1$, $t^* \gg 1/\Delta\mu$ and $\Phi_f(t^*)\gg 1$,
so that by the time $t$ reaches $t^*$ the evolution function has become negligibly small.
In this circumstance the $t^2$ behavior controls the relaxation (for all relevant energy
differences), leading to the Gaussian behavior discussed in the text.
In the non-equilibrium case, the key parameter is therefore $\Gamma^2/\kappa$,
and as this becomes larger than unity, Gaussian behavior results.
In contrast, in the classical (high temperature) Marcus limit,  the role of $\Gamma^2/\kappa$ is replaced by
$2 \alpha k_BT/\omega_c$, while for $k_BT<\omega_c$, the role is played by $\alpha$. If this is small,
one has exponential relaxation, while if this is larger than $1/2$, the kinematics are different and the frequency
dependence is controlled by the bandwidth scale $\omega_c$. Thus the non-equilibrium Marcus-like rate
found here is really a new phenomenon.

We proceed and calculate
the classical Marcus  rate in the high temperature limit.
We substitute Eq. (\ref{eq:phic}) and the short time limit of
Eq. (\ref{eq:phi1Marcus}) into Eq. (\ref{eq:Rbos}),
perform the Fourier transform, and recover the Marcus relation for
the nonadiabatic rate
\beq
\Gamma_b^{\pm}=\left(\frac{\Delta}{2}\right)^2 \sqrt{\frac{\pi}{\lambda k_BT}}e^{-(\mp B +\lambda)^2/(4\lambda k_BT)}.
\label{eq:Gammab}
\eeq
%
The temperature dependence of the rate constant shows an activated
regime for $|\lambda \pm B|\neq 0$, while in the absence of the
barrier, $-B=\lambda$, the rate {\it decreases} with $T$.
The excitation rate $\Gamma_b^{-}$ and the emission rate
$\Gamma_b^+$ are related one to another through an activation
factor as
\beq
\Gamma_b^-/\Gamma_b^{+}=e^{-B/k_BT}.
\eeq
For completeness, we include here other results of the
spin-boson model in the nonadiabatic limit:
The Golden Rule rate to lowest order in $k_BT/\omega_c$ and
$B/\omega_c$ yields ($B>0$) \cite{WeissBook}
\bea
\Gamma_b^{-}&=&\frac{\Delta^2}{4\omega_c} \left(\frac{\omega_c}{2\pi k_BT}\right)^{1-2\alpha}
\nonumber\\
&\times&
\frac{|\tilde \Gamma(\alpha-iB/2\pi k_BT)|^2}{\tilde \Gamma(2\alpha)} e^{-B/2k_BT}.
\eea
Here $\tilde \Gamma(x)$ is the complete Gamma function. For weak
damping, $\alpha \ll 1$, this expression reduces to
 \bea
\Gamma_b^{-}&=&\left(\frac{2\pi k_B
T}{\Delta_{eff}}\right)^{2\alpha} \frac{\pi \alpha
\Delta_{eff}^2}{[(2 \alpha \pi k_B T)^2 +B^2]}
\nonumber\\
&\times&
\frac{B}{e^{B/k_BT}-1},
\label{eq:bosonweak}
\eea
where $\Delta_{eff}$ is an effective tunneling element ($\alpha<1$),
\bea
\Delta_{eff}=\Delta
[\tilde \Gamma(1-2\alpha)\cos(\pi \alpha)]^{\frac{1}{2(1-\alpha)}}
(\Delta/\omega_c)^{\alpha/(1-\alpha)}.
\nonumber\\
\eea
At zero temperature we can calculate the rate exactly for an arbitrary cutoff frequency
($B>0$)
\bea \Gamma_b^{+}(T=0)&=&\frac{\pi\Delta^2}{2\tilde \Gamma(2
\alpha)\omega_c} \left( \frac{B}{\omega_c}\right)^{2\alpha-1}
e^{-B/\omega_c},
\nonumber\\
\Gamma_b^{-}(T=0)&=&0.
\label{eq:bosonT0}
\eea
As expected, at $T=0$ there are transitions only from the upper level to the lower state.

\renewcommand{\theequation}{C\arabic{equation}}
\setcounter{equation}{0}  
\section*{\label{Bosonization}APPENDIX C: Bosonization of the non-equilibrium Fermi edge
Hamiltonian }

We briefly present here some of the relations between
bosonic and fermionic operators, and transform our fermionic
system-bath Hamiltonian into its bosonic analog via bosonization \cite{Giamarchi},
and discuss the issues involved in bosonizing the non-equilibrium version of the model.
For simplicity,
instead of the spin-boson Hamiltonian Eqs.
(\ref{eq:HTotal})-(\ref{eq:Hfermion}), we discuss here the x-ray
edge Hamiltonian (\ref{eq:edge}), describing the interaction of a localized core
hole with two (possibly out-of-equilibrium) metal leads
\bea H&=&\sum_{k,n=L,R}\epsilon_k a_{k,n}^{\dagger}a_{k,n}+
\sum_{k,k',n=L,R}V_{n,n}a_{k,n}^{\dagger}a_{k',n}d^{\dagger}d
\nonumber\\
&+&\sum_{k,k'}\left[ V_{L,R}a_{k,L}^{\dagger} a_{k',R} +
V_{R,L}a_{k,R}^{\dagger} a_{k',L} \right]d^{\dagger}d.
\label{eq:Hfull} \eea
The first term includes two isolated Fermi baths. The second and
third terms describe intra-bath processes (diagonal coupling), and
inter-bath interactions, respectively. $d$ ($d^{\dagger}$) are
annihilation (creation) operators of the core hole. We consider
bands of width $D_0$ and a constant density of states $\rho$.

To solve the equilibrium problem one proceeds as follows.
First, one defines new fermion operators
\begin{eqnarray}
\alpha_{k,+}&=&\cos\frac{\theta}{2}a_{k,R}+\sin\frac{\theta}{2}a_{k,L} \\
\alpha_{k,-}&=&-\sin\frac{\theta}{2}a_{k,R}+\cos\frac{\theta}{2}a_{k,L}
\end{eqnarray}
with
\begin{equation}
\tan\frac{\theta}{2}=\frac{V_{RR}-V_{LL}}{2\sqrt{V_{LR}V_{RL}}}
\end{equation}
in terms of which Eq. (\ref{eq:Hfull}) becomes
\bea H&=&\sum_{k,n=+,-}\epsilon_k \alpha_{k,n}^{\dagger}\alpha_{k,n}+
\sum_{k,k',n=+,-}V_na_{k,n}^{\dagger}a_{k',n}d^{\dagger}d
\label{eq:Hfull2}
\nonumber \\
 \eea
with $V_\pm=\frac{V_{LL}+V_{RR}}{2}\pm\sqrt{\frac{(V_{LL}+V_{RR})^2}{4}+V_{LR}V_{RL}}$.
Crucially, in equilibrium the new fermion variables obey the
usual Fermi statistics
\begin{equation}
\langle \alpha^\dagger_{k,n}\alpha_{k',n'}\rangle =\delta_{k,k'}\delta_{n,n'}f(\varepsilon_k/k_BT),
\label{statistics}
\end{equation}
with $f(\epsilon_k/k_BT)$ as the Fermi-Dirac distribution function.
For this reason, each channel can be bosonized,
leading to the standard result of Schotte and Schotte for
the Fermi edge singularity problem \cite{Schotte}. In a
non-equilibrium situation, while the change of basis can be made, the different
distribution functions for the left and right leads mean that Eq. (\ref{statistics})
does not hold, preventing bosonization in the transformed basis.
One may attempt to proceed by defining the density operator
\bea \rho_n(q)=\sum_{k} a_{k+q,n}^{\dagger} a_{k,n} ,\,\,\,\,
(n=L,R), \eea
which creates particle-hole excitations in the $n$ lead with momentum $q$.
We use it and define bosonic creator and annihilator
operators that obey the bosonic commutation relation
\bea b_{q,n}^{\dagger}&=&\sqrt \frac{2 \pi}{L q}\rho_n(q) \,\,\,\
(q>0),
\nonumber\\
b_{q,n}&=&\sqrt \frac{2 \pi}{L q}\rho_n(-q)\,\,\,\,\,\ (q>0).
\label{eq:boson}
\eea
The distance $L$ is related to the density of states through
$\rho=L/2\pi v_{F}$ with $v_{F}$ as the velocity at the Fermi energy,
taken to be the same for both reservoirs. The fermion field
operators
\bea \Psi_n=\frac{1}{\sqrt L} \sum_ka_{k,n}, \,\,\,\,
\Psi_n^{\dagger}=\frac{1}{\sqrt L}\sum_k a^\dagger_{k,n}, \eea
can be expressed in terms of the boson operators as \cite{Giamarchi}
\beq
\Psi_n= \lim _{\alpha \rightarrow 0} \frac{F_n}{\sqrt{2\pi \alpha}}
\exp \left[ \sum_q \sqrt{ \frac{2\pi} {Lq}}
e^{-\alpha q/2}(b_{q,n}-b_{q,n}^{\dagger}) \right].
\eeq
Here $F_n$ ($F_n^{\dagger}$) are the Klein factors that lower
(raise) the total fermion number in the $n$ reservoir by one. The
chemical potential difference is therefore concealed inside these
factors. $\alpha$ is an arbitrary cutoff that regularizes the
theory and mimics a finite bandwidth. Using these expressions,
the fermionic Hamiltonian
(\ref{eq:Hfull}) translates into a bosonic expression as follows,
\bea
H=H_B +(\Delta_a+\Delta_b) d^{\dagger}d.
\label{eq:Hbosonized}
\eea
$H$ includes the isolated reservoir term
\beq
H_B=\sum_{q,n} v_{F} q b_{q,n}^{\dagger} b_{q,n},
\eeq
and diagonal ($\Delta_a$) and nondiagonal ($\Delta_b$) contributions
\bea \Delta_a&=& \sum_{q,n}V_{n,n} \sqrt {\frac{qL}{2\pi
}}(b_{q,n}^{\dagger}+b_{q,n}),
\nonumber\\
\Delta_b&=&(\tilde V_{L,R} U_LU_R^\dagger  +\tilde V_{R,L}U_RU_L^\dagger ) .
\label{eq:delta}
\eea
Here
\bea U_n&=&F_ne^{-\sum_q \lambda_q(b_{q,n}^{\dagger}- b_{q,n})},
\,\,\, (n=L,R);
\nonumber\\
\lambda_{q}&\equiv& \sqrt{ \frac{2 \pi}{L q}} e^{-\alpha q /2}.
\label{eq:C14}
\eea
All the prefactors are absorbed into the coefficient $\tilde
V_{L,R}= \frac{L}{2\pi \alpha}V_{L,R}$.
Eqs. (\ref{eq:Hbosonized})-(\ref{eq:delta}) reveal that for a
non-equilibrium system bosonization yields a {\it nonlinear}
Hamiltonian with a highly nontrivial form.
Compared with the solution in the fermionic picture (Appendix D),
bosonizing the Hamiltonian does not simplify the calculation,
as it does in equilibrium.
It should be noted, however, that the use of Eqs. (\ref{eq:Hbosonized}), 
(\ref{eq:delta}) and (\ref{eq:C14}) 
yield second cumulant expressions identical to those derived in Appendix D.
\renewcommand{\theequation}{D\arabic{equation}}
\setcounter{equation}{0}  
\section*{\label{Cumulant}APPENDIX D: Derivation of the second cumulant expression}
We derive here the details of the weak coupling correlation
function Eq. (\ref{eq:K2}).
The second cumulant is given by
\beq K_2(t)= -\frac{1}{2} \int_0^t dt_1 \int_0^t dt_2\langle T
F(t_1)F(t_2)\rangle _c, \label{eq:C2} \eeq
where $F=\sum_{k,k',n,n'}V_{n,n'}a_{k,n}^{\dagger}a_{k',n'}$,
$\langle... \rangle_c$ denotes a cumulant average and $T$ denotes
time ordering. For simplicity, we disregard diagonal interactions,
$V_{L,L}=V_{R,R}=0$. The integrand is calculated using Wick's
theorem \cite{Mahan} to yield
\bea C_2(\tau)&\equiv& \langle T F(\tau)F(0)\rangle _c
\nonumber\\
&=&(\rho V_{L,R})^2 \bigg[
\int_{-D_0}^{\mu_L}d \epsilon_1 \int_{\mu_R}^{D_0}d\epsilon_2 e^{i(\epsilon_1-\epsilon_2)\tau}
\nonumber\\
&+& \int_{-D_0}^{\mu_R}d \epsilon_2 \int_{\mu_L}^{D_0}d\epsilon_1
e^{-i(\epsilon_1-\epsilon_2)\tau} \bigg]. \label{eq:intK} \eea
Here $2D_0$ is the bandwidth, $\rho$ is the energy independent
density of states, and $\mu_{L}$ ($\mu_R$) is the chemical
potential at the $L$ ($R$) lead. This expression assumes
zero temperature. Assuming wide bands, $\mu_K<D_0$, the
integrals in (\ref{eq:intK}) can be trivially performed producing
\bea
C_2(\tau) &=&
-2(\rho V_{L,R})^2 \left(\frac{1-e^{-iD_0 \tau}}{\tau}\right)^2 \cos(\Delta\mu \tau)
\nonumber\\
&\overset{D_0\tau \gg 1}{\longrightarrow} & 2(\rho V_{L,R})^2
\frac{D_0^2}{(1+iD_0\tau)^2} \cos(\Delta\mu \tau), \label{eq:Cg} \eea
with the voltage difference $\Delta\mu=\mu_L-\mu_R$. In
equilibrium, the correlation function is therefore given by
\beq C_2^{eq}(t)
=2\frac{(\rho V_{L,R})^2D_0^2}{(1+iD_0t)^2}.
 \eeq
We substitute this expression into Eq. (\ref{eq:C2}), and obtain
the second cumulant approximation for equilibrium situations
\beq
K_2^{eq}(t)=-(\rho V_{L,R})^2\ln(1+D_0^2t^2).
\label{eq:K2eq}
\eeq
This is the standard result for Tomonaga's model  \cite{Schotte}.
In non-equilibrium situations, $\Delta \mu \neq 0$, the
correlation function includes an oscillatory function,
(\ref{eq:Cg}), which can be decomposed into its equilibrium and
non-equilibrium contributions as follows:
\bea C_2^{neq}(t)&=& C_2^{eq}(t)+ C_2^{\Delta\mu}(t); \nonumber\\
C_2^{\Delta\mu}(t)&=&\frac{2 (\rho V_{L,R})^2 D_0^2}{(1+iD_0t)^2} \left[
\cos(\Delta\mu t) -1 \right].
 \eea
The equilibrium term yields Eq. (\ref{eq:K2eq}). We proceed with
the non-equilibrium part. For $\Delta\mu\ll D_0$, $D_0t >1$
\bea
&&K_2^{\Delta\mu}(t)\sim(\rho V_{L,R})^2 \int_0^t dt_1
\int_0^t dt_2 \frac{ \cos(\Delta\mu(t_1-t_2)-1)}{(t_1-t_2)^2}
 \nonumber\\
&&= -(\rho V_{L,R})^2 \Delta\mu \int_0^t dt_1\int_0^t dt_2
\frac{\sin(\Delta\mu(t_1-t_2))}{t_1-t_2} \nonumber\\
&&+  (\rho V_{L,R})^2\int_0^t dt_1  \int_0^{t} dt_2\frac{d}{dt_2}
\left[
\frac{\cos(\Delta \mu(t_1-t_2))-1}{t_1-t_2} \right].
\nonumber\\
 \eea
Exact integration leads to
\bea
K_2^{\Delta\mu}(t)&=&-2(\rho V_{L,R})^2  \left[ \Delta\mu
 t {\rm Si}(\Delta\mu t)- (1-\cos(\Delta\mu t))\right]
\nonumber\\
&+& 2(\rho V_{L,R})^2 \left[ \gamma_e+\ln(\Delta\mu t) -{\rm
Ci}(\Delta\mu t) \right].
\label{eq:K2mu}
\eea
The sine and cosine integrals are defined as ${\rm
Si}(x)=\int_0^{x} \frac{\sin(t)}{t}dt$, ${\rm Ci}(x)=
\gamma_e+\ln(x) +\int_0^{x} \frac{\cos(t)-1}{t}dt$, and
$\gamma_e=0.5772$ is the Euler-Mascheroni constant.
In deriving (\ref{eq:K2mu}) we have used the following identities:
%
$\int {\rm Si}(x) dx= \cos(x) +x {\rm Si}(x)$;
$\int_0^{x} \frac{\sin[(t-\alpha)\beta]}{t-\alpha} dt= {\rm Si}[(x-\alpha)\beta]-
{\rm Si}(\alpha\beta).$
%
The sum of
Eqs. (\ref{eq:K2eq}) and (\ref{eq:K2mu}) is our expression
for the second cumulant (\ref{eq:K2}) with $\nu=\pi \rho V_{L,R}$.
After exponentiating, the first term provides an exponential
relaxation at long times, while the second term yields a power law
contribution.

\renewcommand{\theequation}{E\arabic{equation}}
\setcounter{equation}{0}  
\section*{APPENDIX E: Approximate Analytical Evaluation of Rate Constants}

In this Appendix we present a derivation of the rate constants in the important
limiting cases. We begin from the basic  expression (\ref{eq:Gamapprx})
\begin{equation}
\Gamma_f^+(\omega)=\Re \int_0^\infty
\frac{dt}{\left(iDt\right)^{\beta_{eff}(t)}} e^{i\omega t-A(t)}.
\label{eq:Gamscaled2}
\end{equation}
Here $D$ is an energy scale of the order of the bandwidth, and $\beta_{eff}$ is
an effective exponent which changes from the equilibrium power $\beta$ to
the non-equilibrium power $\beta_{neq}$.
The results of section \ref{Numerics} imply that
\begin{equation}
A(t)=g\phi\left(\frac{t}{t^*}\right),
\label{eq:A}
\end{equation}
with $\phi$ defined such that
$\phi(x\rightarrow 0)\rightarrow x^2$ and $\phi(x\rightarrow \infty) \rightarrow x$. This implies
that the coupling constant $g$ and characteristic time $t^*$ are
\begin{eqnarray}
g&=& \frac{\Gamma^2}{\kappa} \label{gdef}; \\
t^*&=& \frac{\Gamma}{\kappa \Delta \mu}.
\label{eq:tstar}
\end{eqnarray}
In the weak coupling limit, ($\nu \ll 1$), $\Gamma\sim\kappa\sim \nu^2$ so $g\ll1$ and $t^*\sim (\Delta \mu)^{-1}$,
while in the strong coupling limit ($\nu \sim 1$), $\Gamma \rightarrow \infty$ while $\kappa$ saturates, so $g\gg1$ and
$t^* \gg(\Delta \mu)^{-1}$.
We define a dimensionless time coordinate $u=t/t^*$ and frequency
$\omega^*=\omega t^*$ in terms of which the dimensionless relaxation rate $D\Gamma_f^+$ becomes
\begin{equation}
D\Gamma_f^+(\omega)=\Re \int_0^\infty
\frac{du}{\left(iu\right)^{\beta_{eff}(u)}}(Dt^*)^{1-\beta_{eff}(u)}e^{i\omega^*u-g\phi(u)}.
\label{eq:Gamscaled}
\end{equation}
The analysis of the integral in Eq. (\ref{eq:Gamscaled}) requires
some care because the scales which dominate the integral may not
be the scales which dominate the real part of the integral. To
isolate the contributions to the real part, we deform to contour
into the complex plane. Writing $u=x+iy$ we deform the integration
contour into two parts, one running along the imaginary axis
($x=0$) to the point $y=y^*$ at which $i\omega^*-g\partial
\phi/\partial y=0$ and another running parallel to the real axis
along the contour $u=x+iy^*$. Thus we have
\begin{equation}
D\Gamma_f^+(\omega)=I_1+I_2,
\end{equation}
with
\begin{eqnarray}
I_1&=&\Im  \int_0^{y^*}dy \frac{(Dt^*)^{1-\beta_{eff}(y)}}{(-y)^{\beta_{eff}(y)}}e^{-\omega^*y-g\phi(iy)},
\\
I_2&=&e^{-|\omega^* y^*|}\Re\int_0^\infty dx\frac{(Dt^*)^{1-\beta_{eff}(x+iy^*)}}{(ix-y^*)^{\beta_{eff}(x+iy^*)}}
\nonumber \\
&& \hspace{0.7in}\times e^{i\omega^*x-g\phi(x+iy^*)}.
\end{eqnarray}
Here $\Re$ refers to the real part of the integral and $\Im$ to the imaginary part.
We analyze these equations first in weak coupling $g<1$. Let us
begin with $I_1$. Inspection of the second cumulant formula shows
that $y^*<0$ if $\omega<0$. In this case the integral does not
have any imaginary part so for $\omega<0$, $I_1=0$.   Next
consider small positive $\omega$, where we may approximate
$\phi(y)=y^2$ implying
\begin{equation}
y^*=\frac{\omega^*}{2g}=\frac{\omega}{2\Gamma \Delta \mu}.
\label{eq:ystarsmall}
\end{equation}
Thus for $\omega<\Gamma \Delta \mu$ we may use the equilibrium exponent and approximate $\phi=y^2$ obtaining
\begin{equation}
I_1= \Theta(\omega)\sin(\pi \beta )(Dt^*)^{1-\beta}
\int_0^{\frac{\omega}{2\Gamma \Delta \mu}}
\frac{dy}{y^{\beta}}e^{-\omega^*y+gy^2}.
 \label{eq:I1small}
\end{equation}
At the endpoint $\frac{\omega^*}{2g}$ the argument of the
exponential is minimized; the minimum value is
$-\frac{(\omega^*)^2}{4g}=-\frac{\omega^2}{4\kappa\Delta \mu^2}$.
Substituting the maximum  value $\omega=\Gamma \Delta \mu$ and
noting that  in weak coupling $\Gamma \sim \kappa \ll1$ we see
that the argument of the exponential is negligible over the entire
range $\omega<\Gamma \Delta \mu$ and we get
\begin{equation}
I_1=\Theta(\omega)\frac{\sin(\pi \beta )
}{1-\beta}\left(\frac{D\omega}{2\Gamma \Delta
\mu^2}\right)^{1-\beta}.
 \label{eq:I1weaklow}
\end{equation}
For $\omega>\Gamma \Delta \mu$, $y^*>1$, and we  must consider the
form of $\phi$ for large imaginary argument. Inspection of the
second cumulant formula shows that
\begin{equation}
\phi(iy)\sim {\rm cosh} (y) \sim e^y,
\end{equation}
implying $y^* \sim \ln(\omega^*/g)$. In this case the value of the
argument of the exponential at the upper limit of integration is
$-\omega^* \ln(\omega^*/g) \sim -( \omega/\Delta \mu) \ln
[\omega/(\Gamma \Delta \mu)]$. Thus for $\omega \gtrsim
\frac{\Delta \mu}{ \ln 1/\Gamma}$ the upper limit of the integral
may be set to infinity, and for $\omega \gtrsim \Delta \mu$ the
integral is dominated by $y<1$ yielding
\begin{eqnarray}
D\Gamma_f^+(\omega)&=& \Im (-1)^{-\beta} (Dt^*)^{1-\beta}\int_0^\infty\frac{dz}{z^\beta}e^{-\omega^*z} \\
&=& \sin(\pi \beta) \left(\frac{D}{\omega}\right)^{1-\beta}{\tilde
\Gamma}(1-\beta).
 \label{eq:Gamweaklow3}
\end{eqnarray}
Here $\tilde \Gamma(x)$ is the complete Gamma function. Eq.
(\ref{eq:Gamweaklow3}) is simply the usual equilibrium result.
Therefore, in weak coupling, $I_1$ is given by Eq.
(\ref{eq:I1weaklow}) for $\omega \lesssim \Delta
\mu/\ln\Gamma^{-1}$ and by Eq. (\ref{eq:Gamweaklow3}) for $\omega
\gtrsim \Delta \mu$, with a rather broad crossover regime.

We next turn to $I_2$ which is non-zero for both signs of
$\omega$. The frequency regimes are as for $I_1$. For
$\omega<\Delta \mu /\ln\Gamma^{-1}$ the integral is dominated by
large $x$  where $\phi=x$ and the prefactor $e^{-|\omega^*y^*|}$ is
negligible, so that we find
\begin{equation}
I_2\approx(D)^{1-\beta_{neq}}{\tilde \Gamma}(1-\beta_{neq})
\Re\left[\frac{e^{-i\pi \beta_{neq}/2}}{\left(i\omega -\Gamma
\Delta \mu\right)^{1-\beta_{neq}}}\right]. \label{eq:Gamweaklow}
\end{equation}
Note that we have replaced $\beta$ by the long time non-equilibrium
value $\beta_{neq}$.
In the weak coupling limit, $\beta_{neq}\sim \nu^4 \ll1$. Setting
$\beta_{neq}\rightarrow 0$ yields a Lorentzian behavior
\begin{equation}
I_2=\frac{D\Gamma \Delta \mu}{\omega^2+\Gamma^2\Delta\mu^2}.
\label{eq:Gamweaklow2}
\end{equation}
Note that for $\omega>0$ $I_2$ only becomes smaller than $I_1$ for $\omega \sim \Gamma \Delta \mu/\sin \beta \pi\sim\Delta \mu$.
As $\omega$ becomes of the order of $\Delta \mu/\ln\Gamma^{-1}$
the prefactor begins to be important and $I_2$ decays proportional
to $e^{-\frac{\omega}{\Delta\mu} \ln\left(\frac{\omega}{\Delta \mu}\right) }
$ as found by Mitra et al \cite{Aditisemi}.

To summarize, for weak coupling we find a rate which for small
frequencies $\omega<\Delta \mu/\ln\Gamma^{-1}$ is approximately
Lorentzian, with decay constant $\Gamma \Delta \mu$. On the
emission (positive frequency) side, the Lorentzian decay is
overcome by the contribution of $I_1$ and eventually crosses over
the equilibrium rate, Eq. (\ref{eq:Gamweaklow3}), while on the
absorption (negative frequency) side the rate crosses over to
into the $e^{-\frac{\omega }{\Delta\mu}\ln(\frac{\omega}{\Delta \mu})}{\Delta \mu}$ relaxation.

We now take up the strong coupling ($g>1$) limit. For
$\omega<\Gamma \Delta \mu$ Eq. (\ref{eq:I1small}) still applies,
but now  at the endpoint of the integration region the argument of
the exponential can be large. For $\omega^*<\sqrt{g}$ (i.e.
$\omega<\sqrt{\kappa}\Delta \mu$), the variation of the exponent
is not important, and we get
\bea &&I_1^{(A)}= \Theta( \omega)\frac{\sin(\pi
\beta)}{1-\beta}\left( \frac{D\omega}{2\kappa \Delta
\mu^2}\right)^{1-\beta}; \,\,\,\,  |\omega|\ll\sqrt{\kappa}\Delta
\mu.
\nonumber\\
\eea
In contrast, in the opposite high frequency limit we can set the upper
limit of the integration to infinity and drop the $y^2$ term. This
yields
\bea
&&I_1^{(B)}=\Theta(\omega) \sin(\pi \beta ) \left( \frac{D}{ \omega}\right)^{1-\beta} \tilde \Gamma(1-\beta);
\,\,\,\ \omega>\sqrt{\kappa}\Delta \mu,
\label{relaxeqm}\nonumber\\
\eea
which is again the equilibrium ($\Delta\mu=0$) result. This
continues to apply even for $\omega>\Gamma \Delta \mu$. We next
turn to $I_2$. For $\omega<\Gamma \Delta \mu$ we again approximate
$\phi=(x+iy^*)^2$ and find
\begin{equation}
I_2=e^{-\frac{\omega^2}{4\kappa\Delta \mu^2}} (Dt^*)^{1-\beta}
\Re\left[\int_0^\infty \frac{dx}{\left(ix-\frac{\omega}{2\Gamma
\Delta \mu}\right)^\beta}e^{-gx^2}\right].
\end{equation}
The important $x$ are of order $\sqrt{g}$
so that for $\omega<\sqrt{\kappa}\Delta \mu$ we may neglect the $\omega$ in the denominator and get
\bea
&&I_2^{(A)}=
\nonumber\\
&&\cos{\frac{\pi \beta}{2}} e^{-\frac{\omega^2}{4\kappa\Delta \mu}}
\left(\frac{D}{\sqrt{\kappa}\Delta \mu}\right)^{1-\beta} \frac{\tilde \Gamma\left[(1-\beta)/2\right]}{2};
 \,\,\,\,\, | \omega|<\sqrt{\kappa}\Delta \mu.
\nonumber\\
\eea
In the opposite limit $|\omega|>\sqrt{\kappa}\Delta \mu$ we
neglect $x$ in the denominator and find
\bea &&I_2^{(B)}=
\nonumber\\
&&e^{-\frac{\omega^2}{4\kappa\Delta \mu^2}} \cos(\pi
\beta)\frac{D}{2\sqrt{\kappa} \Delta \mu}\left(\frac{2\kappa
\Delta \mu^2}{D\omega}\right)^\beta;
|\omega|>\sqrt{\kappa}\Delta \mu.
\label{marcus11} \nonumber\\
\eea
For the emission rate $\omega>0$, the Marcus rate given by Eq \ref{marcus11}
goes over to the equilibrium power law
behavior, Eq \ref{relaxeqm}, when $I_2^{(B)}$ becomes smaller than $I_1^{(A)}$,
when happens for
$\omega$ slightly larger than $2\sqrt{\kappa}\Delta \mu$.

Finally, if $\omega^*>g$ ($\omega>\Gamma\Delta \mu$) then the
approximation $\phi = y^2$ does not apply and the rate goes over
the $e^{-\omega \ln \omega}$ form discussed above. However, by
this time the rate is so small that this behavior is not relevant.



\begin{thebibliography}{99}
\bibitem{Hamilton}
The Hamiltonian (\ref{eq:Hfermion}) could be also defined using a symmetric
form for the system-bath interaction
$\bar H_{SB}^{(f)}=\sigma_z\sum_{k,n;k',n'}\frac{V_{k,n;k',n'}}{2}a_{k,n}^{\dagger}a_{k',n'}$.
The resulting equilibrium and non-equilibrium phase shifts then relate
to Eqs. (\ref{eq:phaseE})-(\ref{eq:phaseC}) by
$\bar \delta_{\pm}=2\delta_{\pm}(\alpha_1/2, \alpha_2/2,\nu/2)$
and $\bar \delta_{L,R}=2\delta_{L,R}(\alpha_1/2, \alpha_2/2,\nu/2)$.
\bibitem{Marcus-comm}
For strong bias ($\Delta\mu \sim D$) the second cumulant (Appendix D)
yields  at intermediate times
 $K_2(\Delta\mu t\sim 1)=-(\rho V)^2 D \Delta\mu t^2/2$.
For strong coupling, following the discussion in section III.B, we find that
$\Phi_{neq}(t)\sim \kappa D \Delta\mu t^2 $,
see Eq. (\ref{eq:phineqkappa}).
Substituting this result into Eq. (\ref{eq:At})
yields the "normal" Marcus rate with reorganization energy
$\lambda_f= \kappa D$ and driving force $F=\Delta\mu$, in full analogy with the
bosonic system.
\end{thebibliography}
\end{document}